\documentclass[numberedappendix]{emulateapj}

\tightenlines

\countdef\decade=200
\advance\decade by \year
\countdef\hours=201	
\advance\hours by \time
\divide\hours by 60
\countdef\mins=202
\advance\mins by \hours
\multiply\mins by 60
\multiply\hours by 100
\countdef\miltime=203
\advance\miltime by \hours
\advance\miltime by \time
\advance\miltime by -\mins

\newcommand{\ltsim}{\protect\raisebox{-0.5ex}{$\:\stackrel{\textstyle <}
        {\sim}\:$}}
\newcommand{\gtsim}{\protect\raisebox{-0.5ex}{$\:\stackrel{\textstyle >}
        {\sim}\:$}}

\newcommand{\msun}{M_{\odot}}

\newcommand{\Sigmag}{\Sigma_{\rm g}}

\newcommand{\Sigmacomp}{\Sigma_{\rm comp}}

\newcommand{\sigmad}{\sigma_{\rm d}}

\newcommand{\fmol}{f_{\rm H_2}}
\newcommand{\fatm}{f_{\rm HI}}

\newcommand{\calr}{\mathcal{R}}

\newcommand{\rmol}{R_{\rm H_2}}

\newcommand{\phicnm}{\phi_{\rm CNM}}

\newcommand{\Estar}{E^{*}}
\newcommand{\tauR}{\tau_{\rm R}}
\newcommand{\fdissoc}{f_{\rm diss}}

\newcommand{\tauhi}{\tau_{\rm HI}}
\newcommand{\hi}{H\textsc{i} }
\newcommand{\xm}{x_{\rm H_2}}
\newcommand{\Sigmahi}{\Sigma_{\rm HI}}
\newcommand{\Sigmaobs}{\Sigma_{\rm obs}}
\newcommand{\vg}{v_{\rm g}}
\newcommand{\kb}{k_{\rm B}}
\newcommand{\muh}{\mu_{\rm H}}
\newcommand{\taucomp}{\tau_{\rm c}}
\newcommand{\taurd}{\tau_{\rm R\Delta}}
\newcommand{\taucd}{\tau_{\rm c\Delta}}
\newcommand{\fhi}{f_{\rm HI}}
\newcommand{\nc}{n_{\rm CNM}}
\newcommand{\tcnm}{T_{\rm CNM}}
\newcommand{\ttwo}{T_{\rm CNM,2}}
\newcommand{\phimol}{\phi_{\rm mol}}
\newcommand{\sech}{\mbox{sech}}

\defcitealias{krumholz08c}{Paper I}
\defcitealias{blitz06b}{BR06}
\defcitealias{leroy08}{L08}
\defcitealias{wong02}{WB02}
\defcitealias{blitz04}{BR04}
\defcitealias{bigiel08}{B08}

\begin{document}

\title{The Atomic to Molecular Transition in Galaxies.\\
II: HI and H$_2$ Column Densities}

\slugcomment{Accepted to the Astrophysical Journal, October 31, 2008}

\author{Mark R. Krumholz\altaffilmark{1}}
\affil{Department of Astrophysical Sciences, Princeton University, Peyton Hall,
Princeton, NJ 08544 and Department of Astronomy \& Astrophysics, University of California, Santa Cruz, Interdisciplinary Sciences Building, Santa Cruz, CA 95060}
\email{krumholz@ucolick.org}

\author{Christopher F. McKee}
\affil{Departments of Physics and Astronomy, University of California, Berkeley, Campbell Hall,
Berkeley, CA 94720-7304}
\email{cmckee@astro.berkeley.edu}

\author{Jason Tumlinson}
\affil{Space Telescope Science Institute, 3700 San Martin Dr., Baltimore, MD 21218}
\email{tumlinson@stsci.edu}

\altaffiltext{1}{Hubble Fellow}

\begin{abstract}
Gas in galactic disks is collected by gravitational instabilities into giant atomic-molecular complexes, but only the inner, molecular parts of these structures are able to collapse to form stars. Determining what controls the ratio of atomic to molecular hydrogen in complexes is therefore a significant problem in star formation and galactic evolution. In this paper we use the model of H$_2$ formation, dissociation, and shielding developed in the previous paper in this series to make theoretical predictions for atomic to molecular ratios as a function of galactic properties. We find that the molecular fraction in a galaxy is determined primarily by its column density and secondarily by its metallicity, and is to good approximation independent of the strength of the interstellar radiation field. We show that the column of atomic hydrogen required to shield a molecular region against dissociation is $\sim 10$ $\msun$ pc$^{-2}$ at solar metallicity. We compare our model to data from recent surveys of the Milky Way and of nearby galaxies, and show that the both the primary dependence of molecular fraction on column density and the secondary dependence on metallicity that we predict are in good agreement with observed galaxy properties.
\end{abstract}


\section{Introduction}

The formation of molecular hydrogen is a critical step in the transformation of interstellar gas into new stars. The neutral atomic interstellar medium (ISM) in galaxies is generally segregated into cold clouds embedded in a warm inter-cloud medium \citep{mckee77, wolfire03}, and the inner parts of some of these cold atomic clouds harbor regions where the gas is well-shielded against dissociation by the interstellar radiation field (ISRF). In these regions molecules form, and once they do star formation follows.

A full theory of star formation requires as one of its components a method for expressing in terms of observables the fraction of a galaxy's ISM that is in the molecular phase \citep[e.g.][]{krumholz05c}. No models published to date satisfy this requirement, but observations have yielded a number of empirical rules for galaxies' molecular content. Based on \hi and CO mapping of nearby galaxies \citet[hereafter WB02]{wong02} and \citet[hereafter BR04 and BR06]{blitz04,blitz06b} infer that the molecular to atomic surface density ratio $R_{\rm H_2}=\Sigma_{\rm H_2}/\Sigma_{\rm HI}$ in a galaxy varies with the interstellar pressure $P$ needed for hydrostatic balance in the ISM as $R_{\rm H_2}\propto P^{0.92}$, and that the atomic surface density saturates at a maximum value of $\sim 10$ $\msun$ pc$^{-2}$. The observed saturation and a similar dependence of molecular fraction on pressure are also seen in newer surveys such as the HERA CO-Line Extragalactic Survey (HERACLES) that cover a broader range of galaxy properties at higher spatial resolution \citep[hereafter L08]{walter08a, bigiel08, leroy08}. However the physical origin of these patterns is unclear. The samples on which they are based are composed solely of nearby galaxies with a limited range of properties, and in the absence of a physical model it is uncertain how far they can safely be extrapolated to regimes of metallicity, surface density, or other properties not represented in samples of nearby galaxies. 

Theoretical treatments of the problem to date do not yet make such an extrapolation possible. A number of authors have considered the microphysics of H$_2$ formation and the structure of photodissociation regions in varying levels of detail \citep[e.g.][]{vandishoeck86, black87, sternberg88, elmegreen89, draine96, neufeld96, spaans97, hollenbach99, liszt00, liszt02, browning03, allen04}, but none of these treatments address the problem of atomic to molecular ratios on galactic scales.
\citet{wyse86a} and \citet{wang90a, wang90b} present models for cloud formation in galactic disks, but these both rely on prescriptions for the rate of conversion of atomic to molecular gas that are based either on rates of cloud collisions or on Schmidt laws, not on physical models of H$_2$ formation and dissociation. \citet{elmegreen93} gives a theory of the molecular fraction in galaxies that does include a treatment of the H$_2$ formation and self-shielding. However, his model neglects dust shielding, an order unity effect, and it also requires knowledge of a galaxy's ISRF strength, which cannot easily be determined observationally, and its interstellar pressure, which can only be inferred indirectly based on arguments about hydrostatic balance. This makes the model difficult to test or to apply as part of a larger theory of star formation. \citet{schaye04} considers the conditions necessary to form a cold atomic phase of the ISM. The existence of such a phase is a necessary but not sufficient condition for molecule formation, so while \citeauthor{schaye04}'s model provides a minimum condition for star formation, in makes no statements about what fraction of the ISM goes into the molecular phase able to form stars, and thus no statement on the star formation rate that is achieved once the minimum condition is met.

Numerical models are in a similar situation. \citet{hidaka02a} and \citet{pelupessy06} simulate galaxies using subgrid models for H$_2$ formation similar to those presented by \citet{elmegreen93}, and show that they can reproduce some qualitative features of the H$_2$ distribution in galaxies. \citet{robertson08} show that a simulation of a galaxy's ISM that includes radiative heating and cooling in the ionized and atomic phases, coupled with an approximate treatment of H$_2$ formation on grains and dissociation by the ISRF, can reproduce the observed molecular content of galaxies. This suggests that the simulations contain the necessary physical ingredients to explain the observations, but the simulations do not by themselves reveal how these ingredients fit together to produce the observed result. Moreover, like the observed empirical rules, the simulations are based on a very limited range of galaxy properties, and in the absence of a model we can use to understand the origin of the simulation results, it is unclear how to extrapolate. Extending the simulations to cover the full range of galaxy parameters in which we are interested would be prohibitively expensive in terms of both computational and human time.

Our goal in this paper is to remedy this lack of theoretical understanding by providing a first-principles theoretical calculation of the molecular content of a galactic disk in terms of direct observables. In \citet[hereafter Paper I]{krumholz08c}, we lay the groundwork for this treatment by solving the idealized problem of determining the location of the atomic to molecular transition in a uniform spherical gas cloud bathed in a uniform, isotropic dissociating radiation field. In this paper we apply our idealized model to atomic-molecular complexes in galaxies as a way of elucidating the underlying physical processes and parameters that determine the molecular content. We refer readers to Paper I for a full description of our solution to the idealized problem, but here we repeat a central point: for a spherical cloud exposed to an isotropic dissociating radiation field, if we approximate the transition from atomic to molecular as occurring in an infinitely thin shell separating gas that is fully molecular from gas of negligible molecular content, the fraction of a cloud's radius at which this transition occurs is solely a function of two dimensionless numbers:
\begin{eqnarray}
\chi & = & \frac{\fdissoc \sigmad c \Estar_0}{n\calr} \\
\tauR & = & n \sigmad R.
\end{eqnarray}
Here $\fdissoc\approx 0.1$ is the fraction of absorptions of a Lyman-Werner (LW) band photons that produce H$_2$ dissociation rather than simply excitation and radiative decay to a bound state, $\sigmad$ is the dust absorption (not extinction) cross-section per hydrogen nucleus in the LW band, $\Estar_0$ is the free-space number density of LW photons (i.e.\ far outside our cloud), $n$ is the number density of hydrogen nuclei in the atomic shielding layer, $\calr$ is the H$_2$ formation rate coefficient on dust grain surfaces, and $R$ is the cloud radius.

The quantity $\tauR$ is simply a measure of the size of the cloud. It is the dust optical depth that a cloud would have if its density throughout were equal to its density in the atomic region. We may think of $\chi$ as a dimensionless measure of the intensity of the dissociating radiation; formally it is equal to the ratio of the rate at which LW photons are absorbed by dust grains to the rate at which they are absorbed by hydrogen molecules in a parcel of predominantly atomic gas in dissociation equilibrium in free-space. This is a measure of the strength of the radiation field because in strong radiation fields the gas contains very few molecules, so most LW photons are absorbed by dust and $\chi$ is large. In weak radiation fields the gas contains more molecules, which due to their large resonant cross-section dominate the absorption rate, making $\chi$ small. 

Over the remainder of this paper we apply the model of Paper I to atomic-molecular complexes in galaxies. In \S~\ref{giantclouds} we begin by considering giant clouds, which we may approximate as slabs, and in \S~\ref{finite} we extend our treatment to clouds of finite size. In \S~\ref{obscomparison} we compare the model predictions of the previous two sections to observations of atomic and molecular gas. Finally in \S~\ref{conclusion} we summarize and discuss conclusions.

\section{The Atomic Envelopes of Giant Clouds}
\label{giantclouds}

In this section we specialize to the case of giant clouds, which we define as those for which $e^{\tauR}\gg 1$. For these clouds we show in \citetalias{krumholz08c} that the dust optical depth from the cloud surface to the atomic-molecuar transition, $\tauhi$, is a function of $\chi$ alone. We therefore begin our analysis with an estimate of $\chi$.

\subsection{The Normalized Radiation Field}
\label{normrad}

Of the quantities that enter into the normalized radiation field $\chi$, $\fdissoc$ is the most certain, because it is only a very weak function of the spectrum of the dissociating radiation. We therefore take it to have a constant value $\fdissoc\approx 0.1$ independent of environment (\citealt{draine96, browning03}; \citetalias{krumholz08c}). Similarly, $\sigmad$ and $\calr$ are functions of the properties of dust grains. These are both measures of the total surface area of dust grains mixed in the atomic shielding envelope around a molecular cloud; the former measures the area available for absorbing photons, while the latter measures the area available for catalyzing H$_2$ formation. There will of course be an additional dependence of these quantities on the optical and chemical properties of grains, but these effects likely provide only a small fraction of the total variation. To first order, therefore, we expect the ratio $\sigmad/\calr$ to vary little with galactic environment, and we can simply adopt the value from the solar neighborhood. This is
\begin{equation}
\label{sigmar}
\frac{\sigmad}{\calr} = 3.2\times 10^{-5} \frac{\sigma_{\rm d, -21}}{\calr_{-16.5}} \mbox{ s cm}^{-1}
\end{equation}
where $\sigma_{\rm d,-21}=\sigmad/10^{-21}$ cm$^2$, $\calr_{-16.5}=\calr/10^{-16.5}$ cm$^3$ s$^{-1}$, and our best estimates for the solar neighborhood give $\sigma_{\rm d,-21} \approx \calr_{-16.5} \approx 1$ \citep{draine96, wolfire08}.

Unfortunately, $n$ and $\Estar_0$ are considerably harder to determine, since we cannot easily make direct measurements of the atomic density around a molecular cloud or the dissociating radiation field to which it is subjected, particularly for clouds in extragalactic space. (It is possible to determine these quantities for PDRs being produced by individual star clusters -- see \citealt{smith00a} and \citealt{heiner08a,heiner08b} -- but these methods are generally not able to determine mean radiation fields around giant clouds.) However, we can still gain considerable insight into the ratio $\Estar_0/n$ that enters into $\chi$ if we realize that $n$ is not free to assume any value for a given $\Estar_0$. The atomic gas in a galaxy generally comprises regions of both cold and warm gas (cold neutral medium and warm neutral medium, or CNM and WNM, respectively) in approximate pressure balance \citep[e.g.][]{mckee77, wolfire03}. Molecular clouds form in regions where the gas is primarily cold. This is because the the effective opacity to LW photons provided by the small population of molecules found in a given element of predominantly atomic gas varies as $n^2$, so the cold phase, due to its higher density, is far more effective at shielding from LW photons than the warm phase. Thus the $n$ we are concerned with is not the mean density of a galaxy's atomic ISM, it is the density in the cold phase only. In the presumably dense gas in the vicinity of a molecular cloud, most of the mass is likely in the cold rather than the warm phase in any event.

We can estimate the CNM density by using the condition of pressure balance between the cold and warm phases. \citet{wolfire03} show that, for a given ambient FUV radiation intensity $G_0$ (given in units of the \citealt{habing68} field, corresponding to a number density $\Estar_0\approx 4.4 \times 10^{-4}$ LW photons cm$^{-3}$), ionization rate from EUV
radiation and x-rays $\zeta_t$, abundance of dust and polycyclic
aromatic hydrocarbons $Z_d$, and gas phase metal abundance $Z_g$, the
minimum number density $n_{\rm min}$ at which CNM can
exist in pressure balance with WNM is well-approximated by
\begin{equation}
\label{nmin1}
n_{\rm min} \approx 31 G_0' \frac{Z_d'/Z_g'}{1 + 3.1 (G_0'
Z_d'/\zeta_t')^{0.365}}\mbox{ cm}^{-3},
\end{equation}
where the primes denote quantities normalized to their values in the solar
neighborhood. \citet{wolfire03} obtain this expression by constructing a temperature-density relation, determined by balancing the rate of grain photoelectric heating against cooling by the fine structure lines of C\textsc{ii} and O\textsc{i}. Once they have constructed the $T-n$ curve, they identify the temperature at which the pressure is minimized. This is the warmest temperature at which the CNM can be in pressure balance, and thus the corresponding density is the lowest possible CNM density. The primary uncertainty in this expression arises from the abundance, size distribution, and reaction properties of polycyclic aromatic hydrocarbons (PAHs), but changes in PAH properties generally change $n_{\rm min}$ only at the factor of $\sim 2$ level (cf.\ Figure 8 of \citealt{wolfire03}).

In a galaxy where young stars provide the dominant source of radiation
and the IMF is constant, the FUV heating rate and the EUV/x-ray
ionization rate are likely to be proportional to the star formation
rate, and therefore to each other. We therefore assume that $\zeta_t'
= G_0'$. Furthermore, if the physics of dust formation does not vary
strongly from galaxy to galaxy, then the dust and gas phase metal
abundances are likely proportional to the total metallicity $Z$, so we
adopt $Z_d'=Z_g'=Z'$. With these approximations the minimum CNM density becomes
\begin{equation}
\label{nmin}
n_{\rm min} \approx 31 \frac{G_0'}{1 + 3.1 Z'^{0.365}}\mbox{ cm}^{-3}.
\end{equation}
We caution at this point that both the assumptions that $\zeta_t' = G_0'$ and $Z_d'=Z_g'=Z'$ are unlikely to hold in elliptical galaxies, where young stars are not the dominant sources of EUV or x-ray radiation, and where the amount of dust per unit metallicity is known to be different than in spirals. Thus, equation (\ref{nmin}) is unlikely to hold in ellipticals.

The CNM can exist in pressure balance at densities higher than $n_{\rm min}$, so we take the typical CNM density to be 
\begin{equation}
\label{ncnm}
n_{\rm CNM} = \phicnm n_{\rm min}.
\end{equation}
We adopt $\phicnm \approx 3$ as our fiducial value, which gives a CNM density of $n_{\rm CNM}=22$ cm$^{-3}$ and (using \citeauthor{wolfire03}'s $T-n$ relation, equation \ref{wolfiretn} of this paper) a temperature of $T_{\rm CNM}=105$ K, consistent with observations of typical CNM properties in the solar neighborhood. (Near a GMC, we expect $G_0'\sim 10$ rather than $G_0'=1$, due to the proximity of sites of star formation -- Wolfire, Hollenbach, \& McKee 2008, in preparation -- but this does not affect our results, since we only care about the ratio $G_0'/n$.) In practice $\phicnm$ cannot be much larger than this, because pressure balance between the CNM and WNM is possible only over a limited range of CNM densities. If the CNM densities exceeds $n_{\rm min}$ by more than a factor of $\sim 10$ the CNM and WNM again cannot be in pressure balance because it is impossible for the warm phase to have a high enough pressure.

Using equation (\ref{sigmar}) for $\sigmad/\calr$ and equation (\ref{ncnm}) for $n$, and noting that the LW photon number density in the solar neighborhood is roughly $7.5\times 10^{-4}$ cm$^{-3}$ (\citealt{draine78}; \citetalias{krumholz08c}), we find a total estimate for the dimensionless radiation field strength
\begin{equation}
\label{chieqn}
\chi = 2.3 \left(\frac{\sigma_{\rm d,-21}}{\calr_{-16.5}}\right) \frac{1+3.1 Z'^{0.365}}{\phicnm}.
\end{equation}
Note that all explicit dependence on the dust properties, the radiation field, and the atomic gas density have cancelled out of this expression.

Dependence on the dust properties has dropped out for the simple reason explained above:  $\sigmad$ and $\calr$ are both measures of the dust surface area, so their ratio is nearly constant.
We can understand the somewhat more subtle reason that dependence on radiation field and the atomic gas density cancel by examining the physics behind 
expression (\ref{nmin1}). As noted already, the minimum possible density in the cold atomic phase of the ISM corresponds to the density and temperature at which the pressure reaches a local minimum. Because the dependence of the cooling rate on gas temperature is determined almost entirely by the quantum mechanical constants and element abundances that determine the shapes of the C\textsc{ii} and O\textsc{i} cooling curves, and the photoelectric heating rate is essentially independent of temperature, the temperature at which this minimum pressure occurs is nearly fixed at $\approx 240$ K, and does not depend on the background radiation field (c.f.\ equation 34 and Appendix C of \citealt{wolfire03}). Thus the density minimum will simply be the density at which the temperature reaches $\approx 240$ K. Since the heating rate varies as $n \Estar_0$ and the cooling rate as $n^2$, it immediately follows that the density at which a fixed temperature is reached varies as $n\propto \Estar_0$. This explains why $n/\Estar_0$ is nearly constant in the CNM. There is only a weak dependence on metallicity, which arises because the heating rate depends on the charge state of polycyclic aromatic hydrocarbons (PAHs), and this in turn depends weakly on metallicity.

Before moving on, we should note that we have neglected the role of internal radiation in determining where a cloud changes from atomic to molecular. This is justified because most stars that contribute significant amounts of dissociating radiation are born in molecular clouds, but they do not stay internal to those clouds for very long. Most dissociating photons come from massive stars born in clusters that burrow their way out of their parent molecular clouds via their winds and H\textsc{ii} regions in only a few Myr. Thus, most of the dissociating radiation to which a molecular cloud is subjected is delivered externally rather than internally, even if it comes from stars born in that cloud.

This does, however, raise another cautionary point. We have computed the CNM density based on an implicit assumption of pressure balance, and we must consider under what circumstances pressure balance might not hold. One situation in which gas might not reach pressure balance is if it is subjected to hydrodynamic perturbations such as supernova shocks that create rapid and substantial changes in pressure, pushing gas into the unstable regime of density and pressure. Such gas is subject to an instability in which pockets of stable CNM condense within it, leaving behind a lower density ambient medium that expands to become stable WNM \citep[e.g.][]{audit05a}. Thus the typical gas density will be significantly different than the value $n_{\rm CNM}$ that we have estimated only if the time between successive shocks that drive gas into instability is small compared to the time required for this instability to operate. This is of order the cooling time scale, which \citet{wolfire03} estimate to be
\begin{equation}
t_{\rm cool} \approx 7.7 \left(\frac{T}{10^4\mbox{ K}}\right)^{1.2} \left(\frac{nT}{3000\mbox{ K cm}^{-3}}\right)^{-0.8}\mbox{ Myr},
\end{equation}
where $n$ and $T$ are the gas number density and temperature. For typical CNM conditions near the Solar circle this is $\sim 40$ kyr, while for typical WNM conditions it is slightly under 10 Myr. Higher values of $G_0'$, as are expected near GMCs, reduce these to a few kyr and a few Myr, respectively. Thus perturbations that produce velocities $\ltsim 5$ km s$^{-1}$, i.e.\ not fast enough to induce shocks in the WNM, are essentially ineffective at driving the gas out of equilibrium, since any CNM gas they disturb will re-equilibrate very quickly. Stronger perturbations such as supernova blast waves that drive WNM unstable can keep substantial amounts of gas out of pressure balance only if they recur on time scales of a few Myr or less. For both supernova blast waves and the shocks induced by the ubiquitous turbulence in the atomic ISM the recurrence time is $\sim 10$ Myr, larger (although not hugely so) than the equilibration time \citep{wolfire03}. We can therefore conclude that the typical atomic envelope around a molecular cloud is likely to be close to pressure balance between CNM and WNM. Some envelopes will have been subjected to a strong shock in the last few Myr, and these may have CNM densities substantially different than $n_{\rm CNM}$, but they will be in the minority.

Alternately, one could consider a galaxy in which the average galactic environment is so extreme that no two-phase equilibrium is possible, i.e.\ the pressure is so low that only WNM is stable, or the pressure is so high that only CNM is stable. \citeauthor{wolfire03}\ and \citet{schaye04} both find that a pressure so low that only WNM exists is consistent with vertical hydrostatic balance only in the very diffuse outer parts of galactic disks. For the Milky Way, \citeauthor{wolfire03}\ find that a CNM phase can exist everywhere the mean ISM density is $\gtsim 0.2$ cm$^{-3}$, which is everywhere in the Milky Way inside $\sim 15$ kpc in Galactocentric radius; \citeauthor{schaye04} estimates that hydrostatic balance requires the existence of a CNM phase any time the local gas surface density exceeds $3-10$ $\msun$ pc$^{-2}$, which is true over a similar region. Thus, we can conclude that a pure WNM is unlikely to exist anywhere except in the far outer regions of galactic disks; in such regions, equation (\ref{chieqn}) underestimates the dimensionless radiation field strength, and thus we will overestimate the molecular fraction. The converse possibility is a galaxy in which the pressure is so high that no WNM is present, only CNM; however, \citeauthor{wolfire03}\ find that that WNM can be present any time the mean ISM density $\ltsim 70$ cm$^{-3}$, unless the CNM is entirely confined by a surrounding hot ionized medium. The case where no WNM exists because $n\gtsim 70$ cm$^{-3}$ corresponds to an entire galaxy whose mean density matches that of a typical molecular cloud in the Milky Way, and such conditions are only found in starburst systems where the molecular fraction is essentially unity. In such cases our models will provide only an upper limit to the amount of \hi present, but even then our model may apply if the atomic shielding layer occurs far from the galactic midplane where the density is lower. The latter possibility, of an ISM consisting solely of cold atomic gas and hot ionized gas, appears not be be realized in nature. Thus, in summary, we expect our two-phase model to apply everywhere in galaxies except in their far outer parts, where the density is so low that no cold phase exists, and at the midplanes of starburst galaxies, where the density is so high that no warm phase can exist.

\subsection{The Shielding Column}
\label{shieldcolumn}

The normalized radiation field $\chi$ is the primary factor controlling the size of the atomic gas column that is needed to shield a molecular cloud against dissociation, and thus in determining fraction of the gas in a galaxy is molecular. This quantity in turn determines what fraction of a galaxy's ISM is molecular, and therefore available for star formation, and what fraction is atomic, and thus without star formation. The invariance of $\chi$ across galactic environments has important consequences for atomic-molecular complexes. First, $\chi$ measures the relative importance of dust shielding and H$_2$ self-shielding, so our results that $\chi\sim 1$ across galaxies implies that dust- and self-shielding contribute nearly equally in essentially all galactic environments. We do not expect to find clouds where either dust- or molecular-shielding completely dominate except near strong sources of dissociating radiation where the atomic ISM is not in pressure balance or in regions of such low or high density that the atomic ISM does not have two phases.

\begin{figure}
\plotone{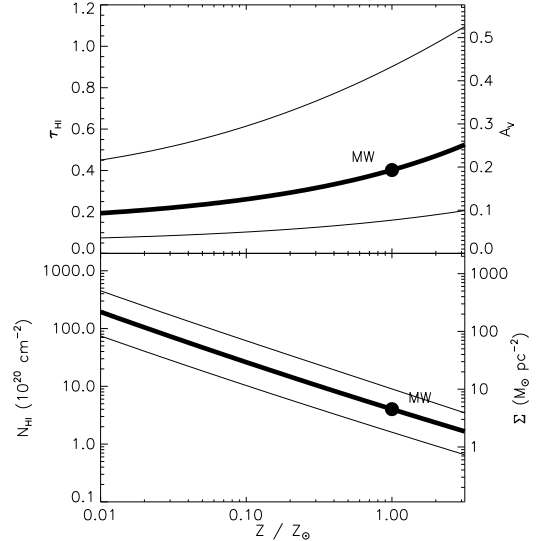}
\caption{
\label{taushieldlarge}
The dust optical depth of the \hi shielding layer $\tauhi$ (\textit{upper panel}), and the corresponding \hi column $N_{\rm \hi}$ (\textit{lower panel}) for very large clouds, $\tauR\rightarrow\infty$. We show these results computed for $\phicnm=3$ (\textit{thick lines}) and for $\phicnm=1$ and $10$ (\textit{upper and lower thin lines, respectively}). The circles indicate our values for the Milky Way, $Z'=1$, for our fiducial $\phicnm=3$: $\tauhi=0.40$, $N_{\rm HI}=4.0\times 10^{20}$. To compute $N_{\rm HI}$ from $\tauhi$, we assume a dust absorption cross section per H nucleus in the LW band of $\sigmad=10^{-21} Z/Z_{\odot}$. We also show the visual extinction $A_V$ corresponding to our $\tauhi$ and the mass column density $\Sigma$ corresponding to $N_{\rm \hi}$. For the former we have assumed $A_V/\tauhi=0.48$, following the models of \citet{draine03a, draine03b, draine03c} as explained in the text. To compute the latter we assume a mean particle mass per H nucleus of $2.34\times 10^{-24}$ g cm$^{-2}$, corresponding to a standard cosmic mixture of H and He. 
}
\end{figure}

Second, the dust optical depth through the \hi shielding layer, $\tauhi$, is primarily a function of $\chi$; the dependence on $\tauR$ arises from geometric effects, and is greater than order unity only if the molecular region inside a cloud is a small fraction of its size. (Strictly speaking this is the optical depth of the CNM, not of all the atomic gas. As noted above most of the atomic gas around a molecular cloud is probably CNM, but there could be significant amounts of WNM along the line of sight that is not associated with that cloud, and which does not contribute to shielding it.) This implies that the dust optical depth through the atomic envelopes of molecular clouds should be roughly constant across galactic environments, at least as long as this optical depth is not close to that of the entire atomic-molecular complex. The only significant variation will be a weak increase in $\tauhi$ with metallicity. Correspondingly, the total \hi gas column will decrease with metallicity to a power less than unity, since $\tauhi$ increases slightly with metallicity, but the column of \hi required to achieve this dust optical depth decreases with metallicity. To illustrate for these effects we solve for $\tauhi$ as a function of $Z$ in the limit $\tauR\rightarrow\infty$ using the formalism of \citetalias{krumholz08c}, and plot the result in Figure \ref{taushieldlarge}. 

As the plot shows, four our fiducial model $\phicnm=3$, we predict that the \hi layer around a molecular cloud in the Milky Way, $Z'=1$, should have an absorption optical depth of $\tauhi=0.40$ to LW photons, corresponding to an \hi column $N_{\rm \hi}=4.0\times 10^{20}$ cm$^{-2}$ (mass column density $\Sigma=4.5$ $\msun$ pc$^{-2}$), assuming a dust absorption cross section per H nucleus of $\sigmad=10^{-21}$ cm$^{-2}$. It is important to note that all of these values represent the absorption column on \textit{one side} of a giant cloud. A 21-cm observation would detect the shielding column on both sides for a cloud exposed to the ISRF on both sides, so the detected \hi column would be \textit{double} the values given in Figure \ref{taushieldlarge}. As a shall see in \S~\ref{finite}, the column is somewhat larger for a cloud of finite size.

It is important to note that the shielding column we have calculated is somewhat different than the atomic-to-molecular transition column density reported for the Milky Way by \citet[$\log N(H)=20.7$]{savage77} and for the LMC and SMC by \citet[$\log N(H)\ge 21.3$ and $\ge 22$, respectively]{tumlinson02a}. These values are the total column densities along pencil-beam lines of sight at which the fraction of the gas column in the form of H$_2$ reaches about $10\%$ of the total. In contrast, in the two-zone approximation we adopt in \citetalias{krumholz08c}, we assume that the atomic-to-molecular transition is infinitely sharp, and under this approximation the shielding column we report is the column at which the gas goes from fully atomic to fully molecular. Were the transition truly infinitely sharp as we have approximated it to be, the ratio of H$_2$ to total column density would be zero at our computed shielding column. Comparing our theoretical shielding columns to the detailed numerical radiative transfer models we present in \citetalias{krumholz08c} shows that in reality, for conditions typical of the Milky Way, the ratio of H$_2$ column to total column at our calculated transition column $N_{\rm HI}$ is roughly 20\%. Since this is a factor of 2 larger than the 10\% ratio used in the observationally-defined transition column, and in our simple model the H$_2$ fraction increases linearly with total column density once we pass our predicted transition point, we expect our shielding column to be a factor of $\sim 2$ larger than the values reported by \citeauthor{savage77}\ and \citeauthor{tumlinson02a} We compare our model predictions to these data sets in more detail in \S~\ref{galobs}.

Using the extinction and absorption curves of \citet{draine03a,draine03b,draine03c}, the ratio of visual extinction to 1000 \AA\ absorption is $A_V/\tauhi=0.48$ for \citeauthor{draine03a}'s $R_V=4.0$ model, so the visual extinction corresponding to $\tauhi=0.40$ is $A_V=0.19$. Adopting the $R_V=5.5$ curve instead, appropriate for denser clouds, gives $A_V=0.28$, while $R_V=3.1$, for diffuse regions, gives $A_V=0.13$. Our estimates for the LW dust optical depth and visual extinction vary little with metallicity, changing by only a factor of $2.7$ for a metallicity ranging from $10^{-2} Z_{\odot}$ to $10^{0.5} Z_{\odot}$.

The variation between the curves with $\phicnm=1,3,10$ show the full plausible range of variation in shielding column arising from our uncertainty about the true density in the atomic envelopes of molecular clouds. The $\phicnm=1$ and $10$ curves are both within a factor of $2.6$ of the fiducial model, so this is an upper bound on our uncertainty. The actual error is likely to be smaller than this, since $\phicnm=1$ and $10$ correspond to the extreme assumptions that the CNM assumes is minimum or maximum possible equilibrium densities.

We can obtain a quick approximation to the results shown in Figure \ref{taushieldlarge} simply by noting that at solar metallicity our fiducial normalized radiation field is $\chi=3.1$, and we show in \citetalias{krumholz08c} that for a giant cloud with $\chi<4.1$ (corresponding to $Z'<2.5$ for our fiducial parameters) the LW dust optical depth through the atomic shielding layer is 
\begin{equation}
\tauhi = \frac{\psi}{4},
\end{equation}
where 
\begin{equation}
\label{psidefn}
\psi=\chi\frac{2.5+\chi}{2.5+\chi e}.
\end{equation}
The dust-adjusted radiation field $\psi$ is a function only of metallicity; for our fiducial parameters $\phicnm=3$ and $\sigma_{\rm d,-21}/\calr_{-16.5}=1$, and Milky Way metallicity, $Z'=1$, we obtain $\psi=1.6$. Moreover, the dependence on metallicity is weak: at $Z'=1/10$, $\psi=1.0$, while at $Z'=1/100$, $\psi=0.77$. Because $\psi$ depends only on metallicity, we can also express the characteristic \hi shielding column on one side of a giant cloud solely as a function of $Z'$:
\begin{eqnarray}
\label{sigmalarge}
\Sigma_{\rm \hi} & = & \frac{\muh}{\sigmad} \tauhi(Z',\phi_{\chi}) \\
& = & 4.5\,\msun\mbox{ pc}^{-2} \frac{f(Z', \phi_{\chi})}{\sigma_{\rm 0,-21} Z'}
\end{eqnarray}
where $\muh=2.34 \times 10^{-24}$ g is the mean mass per hydrogen nucleus and $\sigma_{\rm 0,-21}$ is the dust absorption cross-section at Milky Way metallicity ($\log Z'=0$) in units of $10^{-21}$ cm$^2$. The function $f(Z',\phi_{\chi})$ is given by
\begin{eqnarray}
f(Z',\phi_{\chi}) & = & 0.54 \left(0.32+Z'^{0.385}\right) \phi_{\chi}^{-1} \cdot {}
\nonumber \\
& & \qquad
\left(\frac{1.05 \phi_{\chi}+0.42+Z'^{0.385}}{0.39\phi_{\chi}+0.42+Z'^{0.385}}\right),
\end{eqnarray}
where
\begin{equation}
\phi_{\chi} \equiv \left(\frac{\phicnm}{3}\right) \left( \frac{\calr_{-16.5}}{\sigma_{\rm d,-21}}\right),
\end{equation}
$f(1,1)=1$, and for our fiducial parameters $\phi_{\chi} = 1$. The numerical factors that appear in $f(Z',\phi_{\chi})$ are derived simply by substituting equation (\ref{chieqn}) for $\chi$ into equation (\ref{psidefn}) and thence into equation (\ref{sigmalarge}). These equations, and therefore the numerical values in the function $f$, depend solely on microphysical constants that describe the properties of molecular hydrogen and the chemistry of its formation on grain surfaces (which set $\fdissoc$ and $\sigmad/\calr$) and the shapes of the C\textsc{ii} and O\textsc{i} cooling curves (which set the ratio $\Estar_0/n$). We have therefore calculated the shielding column to good approximation solely in terms of microphysical constants.

\section{The Atomic Envelopes of Finite Clouds}
\label{finite}

\subsection{Formulation of the Problem}

To account for the fact that clouds have finite sizes and column densities, and that these can be quite small in dwarf galaxies or other low-pressure environments, we must examine the second dimensionless number that characterizes H$_2$ formation and shielding: $\tauR=n\sigmad R$. Consider a cloud of known, fixed column density $\Sigmacomp$. If atomic-molecular complexes were of uniform density then we could find $\tauR$ simply by multiplying $\Sigmacomp$ by the dust cross section per unit mass $\sigmad/\muh$, where $\muh\approx 2.34\times 10^{-24}$ g is the mean mass per H nucleus. However, the atomic region is warmer and has a lower mean mass per particle than the molecular one, and thus has a correspondingly lower density. This reduces the dust optical depth through it. Since it is the density and dust optical depth through the atomic shielding layer that matters, we must estimate $\tauR$ using the value of $n$ appropriate for the atomic gas rather than the mean density in the complex. In other words, the quantity we want is
\begin{equation}
\tauR=\nc\sigmad R,
\end{equation}
and we define
\begin{equation}
\phimol \equiv \frac{n_{\rm mol}}{\nc}
\end{equation}
as the ratio of densities. Here $n_{\rm mol}$ and $\nc$ are the number densities of hydrogen nuclei in the molecular and CNM phases of the ISM, respectively.

In the Milky Way, typical molecular cloud densities are $n_{\rm mol} \approx 100$ cm$^{-3}$ \citep{mckee07b}, while observations of the giant \hi clouds around these molecular regions find typical densities $\nc\approx 10$ cm$^{-3}$ \citep{elmegreen87}, suggesting $\phimol\approx 10$. We do not expect this ratio to vary strongly between galaxies, so we should generally find $\phimol\approx 10$.

We can can make an independent argument for $\phimol\approx 10$ by considering thermal pressure balance across the atomic-molecular interface. This argument only applies to gas near an atomic-molecular transition surface, which may or may not include the bulk of the gas in a cloud, but it does provide an estimate for the density ratio near the interface. Pressure balance requires that
\begin{equation}
\label{phimoleqn}
\phimol = 1.8 \frac{T_{\rm CNM}}{T_{\rm mol}},
\end{equation}
where $T_{\rm mol}$ and $T_{\rm CNM}$ are the temperatures in the molecular and cold neutral atomic media, respectively, and the factor of $1.8$ accounts for the difference in mean number of particles per H nucleus in the two phases. Across a very wide range of galactic environments the temperature in the molecular phase of the ISM is $T_{\rm mol} \approx 10-20$ K, as a result of the balance between grain photoelectric heating and CO cooling. Since are interested in gas at the edge of the molecular region, we adopt $T_{\rm mol} = 20$ K as typical. (Temperatures are somewhat higher in starburst systems, but in these galaxies the molecular fraction is essentially unity in any event.) Using the model of \citet{wolfire03} for the atomic medium, and making the approximations $Z'_d=Z'_g=Z'$ and $\zeta'_t=G'_0$ as in \S~\ref{normrad}, the density-temperature relation in the atomic gas is
\begin{equation}
\label{wolfiretn}
\nc \approx \frac{20\, G_0' \ttwo^{-0.2} e^{1.5/\ttwo}}{1+2.6 (\ttwo^{1/2} Z')^{0.365}},
\end{equation}
where $\ttwo=\tcnm/(100\mbox{ K})$. Combining this with equations (\ref{nmin}) and (\ref{ncnm}) enables us to write an implicit equation for the CNM temperature in terms of $\phicnm$ and $Z'$:
\begin{equation}
\label{tcnmeqn}
\frac{20 \, \ttwo^{-0.2} e^{1.5/\ttwo}}{1+2.6 (\ttwo^{1/2} Z')^{0.365}} = \phicnm\frac{31}{1+3.1Z'^{0.365}}.
\end{equation}
Substituting the solution to this equation in equation (\ref{phimoleqn}) immediately gives us $\phimol$, the ratio of the number densities of H nuclei in the molecular and CNM gas. For our fiducial $\phicnm=3$, we find $\phimol=9.6$ at $Z'=1$, varying by only a few percent for metallicities in the range $Z'=10^{-2}-10^{1}$. Given the encouraging agreement between this and the value $\phimol\approx 10$ we find observationally, we adopt the value of $\phimol$ given by equations (\ref{phimoleqn}) and (\ref{tcnmeqn}) as our standard value for the remainder of this work.

A spherical cloud that consists of a molecular core of number density $n_{\rm mol}$ and an outer atomic envelope of number density $n_{\rm CNM}$ has a mean column density
\begin{equation}
\label{sigmacompeq}
\Sigmacomp = \frac{4}{3} \muh n_{\rm CNM} R \left[1+\left(\phimol-1\right)\xm^3\right],
\end{equation}
where $\xm$ is the fraction of the cloud radius at which it transitions from molecular to atomic, i.e.\ $\xm=1$ corresponds to a cloud that is molecular throughout and $\xm=0$ to one that is atomic throughout (see Paper I). It is convenient to rewrite this in terms of an optical depth
\begin{eqnarray}
\label{taucompdefn}
\taucomp & \equiv & \frac{3}{4}\left(\frac{\Sigmacomp\sigmad}{\muh}\right) \\
& \rightarrow & 0.067\, Z' \Sigma_{\rm comp,0},
\end{eqnarray}
where $\Sigma_{\rm comp,0}=\Sigmacomp/(1\,\msun\mbox{ pc}^{-2})$ and the arrow in the second step indicates that we have used our fiducial $\sigmad=10^{-21} Z'$ cm$^2$. Equations (\ref{sigmacompeq}) and (\ref{taucompdefn}) imply
\begin{equation}
\label{xmtaur}
\taucomp = \tauR \left[1+\left(\phimol-1\right) \xm^3\right].
\end{equation}
Note that neither $\tauR$ nor $\taucomp$ is the true center-to-edge dust optical depth of the complex; 
$\tauR$ is the optical depth the complex would have if its density were $n_{\rm CNM}$ throughout, and 
$\taucomp$ is the optical depth it would have if its atomic and molecular gas were mixed uniformly rather than spatially segregated.

We are now in a position to compute the shielding column and the atomic and molecular fractions in finite clouds. If we consider a complex of a given column density $\Sigmacomp$ and metallicity $Z'$, and we take the dust opacity to be given by its Milky Way value adjusted for metallicity, $\sigmad = \sigma_{\rm d,MW} Z'$ with $\sigma_{\rm d,MW}=10^{-21}$ cm$^{-2}$, then equations (\ref{taucompdefn}) and (\ref{xmtaur}) give one constraint on the unknowns $\tauR$ and $\xm$ from pressure balance between the atomic and molecular phases. Dissociation-formation equilibrium, as computed in \citetalias{krumholz08c}, gives a second constraint. We show in \citetalias{krumholz08c} how to compute the value of $\xm$ for a given $\tauR$ and $\chi$: its value is given implicitly by the solution to equations (33) and (37), or (43) and (44), of that paper. Since we have already computed $\chi$ in terms of the metallicity (equation \ref{chieqn}), a choice of $\Sigmacomp$ and $Z'$ fully determine the two unknowns $\tauR$ and $\xm$.

\subsection{Numerical Solution}

\begin{figure}
\plotone{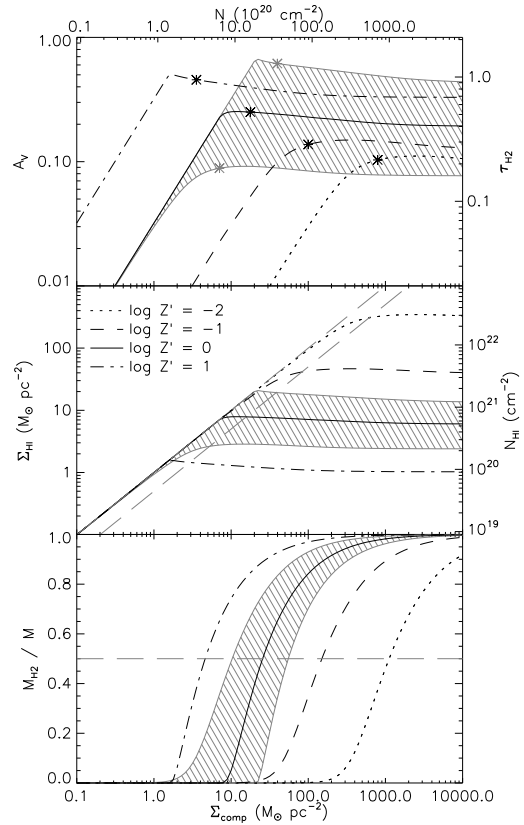}
\caption{
\label{finitecloud}
Visual extinction $A_V$ (\textit{top panel}), \hi column density $\Sigmahi$ (\textit{middle panel}), and H$_2$ fraction $M_{\rm H_2}/M$ (\textit{bottom panel}) in finite cloud complexes, as a function of complex mass column density $\Sigmacomp$ (or number column density $N$). In each panel the curves shown are for metallicities $Z'$ running from $10^{-2}-10^1$, as indicated. The hatched regions centered on the $Z'=0$ curve indicate the range of models with $\phicnm=1-10$, while all other curves are for our fiducial value $\phicnm=3$. In the top panel, the asterisks on the curves indicate the value of $\Sigmacomp$ for which the molecular fraction falls to $1/2$. The curves to the left of this point should be regarded as lower limits on the visual extinction to the molecular region. In the middle panel the two parallel dashed lines indicate ratios of $M_{\rm HI}/M=1/2$ and $1$. Since values of $M_{\rm HI}$ below $1/2$ are lower limits, once our models curves cross the lower of these lines, they true solution could be anywhere between them. Similarly, the dashed line in the bottom panel corresponds to $M_{\rm H_2}/M=1/2$, and when the model curves fall below this they should be treated as upper limits.
}
\end{figure}

We can either solve this system of nonlinear algebraic equations numerically, or approximate the solution analytically. We first show the results of a numerical calculation for a variety of values of $\Sigmacomp$ and $Z'$ in Figure \ref{finitecloud}. Rather than giving $\tauR$ and $\xm$ directly, which are not particularly interesting because we cannot measure them directly, we plot three derived quantities of interest which in at least some circumstances we can observe: the LW optical depth and visual extinction from the cloud surface to the atomic-molecular transition surface along a radial path, the \hi column averaged over the entire cloud, and the total H$_2$ mass fraction over the entire cloud.

It is important to point out that the value of $A_V$ we report is measured differently than the \hi column density $\Sigma_{\rm HI}$, or than the column densities we will use in \S~\ref{obscomparison}. The column density is measured by averaging the mass per unit area over the entire complex, while $A_V$ is measured along a single pencil beam from the surface of the cloud to the atomic-molecular transition along a radial trajectory. The former quantity is more analogous to what is measured in an observation using a telescope beam that only marginally resolves or does not resolve a complex, while the latter is more closely analogous to a measurement of the extinction of a background point source through a cloud. We also caution that, for reasons we discuss in \S~\ref{geomerror}, our predictions are only accurate for molecular mass fractions $>1/2$. (This is in the worst case of very low metallicity and intermediate $\phicnm$; our accuracy range expands as metallicity increases toward solar and as $\phicnm$ gets smaller or larger than $3$.) Below this limit our calculations yield only upper limits on the molecular fraction, not firm predictions. This confidence limit is shown in the Figure \ref{finitecloud}.

The plots immediately yield a number of interesting results. First, our prediction of nearly constant $A_V$ through the atomic shielding layers around molecular clouds continues to hold whenever there is a significant molecular fraction, even for finite clouds. Our prediction of a characteristic $A_V\approx 0.2$ through atomic shielding envelopes of molecular clouds therefore continues to apply.

We also find that there is a saturation in the \hi column density at roughly $6$ $\msun$ pc$^{-2}$ for solar metallicity, which rises by a factor of a somewhat less then ten for every decade by which the metallicity declines. The \hi column saturates simply because once $\Sigmacomp$ is large enough, the cloud column densities become so large that they are effectively in the infinite cloud limit. At this point the shielding column is geometry-independent, and is determined solely by the normalized strength of the radiation field, $\chi$, a value that does not vary much from galaxy to galaxy. Once $\Sigmacomp$ is sufficiently large to put a complex in the large cloud limit, adding additional mass simply increases the size of the shielded molecular layer, so the H$_2$ fraction just rises smoothly. The saturation value of $6$ $\msun$ pc$^{-2}$ at solar metallicity is set by a combination of the fundamental constants describing H$_2$ formation and dissociation, the shape of the C\textsc{ii} and O\textsc{i} cooling curves (which determine the CNM density and temperature), and the properties of interstellar dust grains, which set the ratio $\sigmad/\calr$.

\subsection{Geometric Uncertainties for Finite Clouds}
\label{geomerror}

\begin{figure}
\plotone{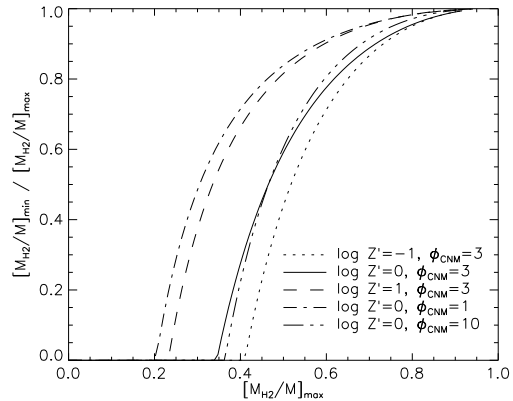}
\caption{\label{hocomparison}
Ratio of minimum to maximum predicted H$_2$ mass fraction versus maximum predicted H$_2$ mass fraction, for a variety of values of $Z'$ and $\phicnm$, as indicated. Curves for $\log Z' < -1$ are not shown because they are indistinguishable from those for $\log Z' = -1$.
}
\end{figure}

In \S~4.7 of \citetalias{krumholz08c} we show that our method for determining the molecular abundance in finite clouds suffers from a systematic uncertainty arising from our imperfect knowledge of the opacity along rays that pass through the atomic envelope of a cloud. For our fiducial model we take the opacity along these rays due to molecules mixed into the atomic gas to be set by the value of the dissociation radiation field at the surface of the zone where molecules dominate the opacity. Our results depend on this approximation very little except at low molecular volume fraction, $\xm^3 \ltsim 0.2$; in that case the uncertainty about this approximation means that our model enables us to predict only an upper limit on the molecular fraction, not an exact value.

In this paper we are concerned with molecular mass rather than volume fractions, so we must quantify that uncertainty. To do so, we proceed as in \S~4.7 of \citetalias{krumholz08c}: we adopt the opposite assumption, that opacity along rays passing through the region of the cloud where molecules dominate the opacity but still constitute a small fraction of all H nuclei is infinite. We then repeat the calculations of \S~\ref{finite} following this assumption: for a given $\Sigmacomp$ and $Z'$, rather than solve the system of equations formed by equation (\ref{xmtaur}) of this paper and equations (33) and (37) or (43) and (44) of \citetalias{krumholz08c}, we instead solve equation (\ref{xmtaur}) together with equations (69) and (70) or (71) of \citetalias{krumholz08c}. Doing so gives a lower bound on the molecular content for a given $\Sigmacomp$ and $Z'$. By comparing the results in this case to our fiducial calculations as presented in \S~\ref{finite}, we obtain an estimate of the uncertainty of our results.

In Figure \ref{hocomparison} we show the results of this exercise. On the $y$-axis we show the ratio of the H$_2$ mass fraction predicted using our maximum opacity assumption, which represents the minimum possible molecular content, divided by the value produced by our fiducial model, which represents the maximum. This gives an estimate of our uncertainty. The $x$-axis indicates the H$_2$ mass fraction predicted using the fiducial assumption we make elsewhere in the paper. As the plot shows, the two calculations differ most at low $Z'$ and intermediate $\phicnm$. In this case the calculations differ by a factor of a few for H$_2$ mass fractions below $\sim 0.5$. If we adopt a factor of $3$ as an accuracy goal, this means that for cases where we predict an H$_2$ mass fraction below $0.5$, and at low metallicity, our predictions should be taken only as upper limits. For solar metallicity or higher our confidence range extends down to molecular mass fractions around $0.4$, and we attain upper limits below this.

\subsection{Analytic Approximation}

We can gain additional insight into the behavior of the solution by constructing an analytic approximation. The ratio of the molecular mass $M_{\rm H_2}$ to the total complex mass $M$ is
\begin{equation}
\label{molfrac}
\fmol \equiv \frac{M_{\rm H_2}}{M} = \frac{\phimol \xm^3}{1+\left(\phimol-1\right)\xm^3}.
\end{equation}
We wish to obtain an approximation for this in terms of the known quantities $\psi$ (given in terms of metallicity by equations \ref{chieqn} and \ref{psidefn}) and $\taucomp$ (given in terms of complex column density by equation \ref{taucompdefn}). We show in \citetalias{krumholz08c} that for $\psi\ltsim 3$ and molecular volume fractions $\xm^3 \gtsim 0.15$, a range in parameter space that includes most of our models for realistic cloud parameters, the molecular volume is well-approximated by
\begin{equation}
\label{xmapprox}
\xm^3 \approx 1 - \frac{3\psi}{4\taurd},
\end{equation}
where for convenience we have defined
\begin{equation}
\taurd \equiv \tauR + a\psi,
\end{equation}
and $a=0.2$ is a numerical parameter that is optimized for agreement between the approximate and numerical solutions. Substituting this approximation into the condition for pressure balance, equation (\ref{xmtaur}), gives
\begin{equation}
\label{taucapproxeqn}
\taucomp = \tauR \left[\phimol+\frac{3\psi}{4\taurd}\left(1-\phimol\right)\right].
\end{equation}
As in \citetalias{krumholz08c}, our approach to obtaining an analytic solution is to perform a series expansion in $a$. We therefore define
\begin{equation}
\taucd \equiv \taucomp \left(1+\frac{a\psi}{\tauR}\right),
\end{equation}
so that $\taucd/\taucomp = \taurd/\tauR$. Using this definition of $\taucd$ together with equation (\ref{taucapproxeqn}) for $\taucomp$ implies that
\begin{equation}
\label{taucdeqn}
\taucd = \phimol \left(\taurd - \frac{3}{4}\psi\right) + \frac{3}{4}\psi.
\end{equation}
If we now use our approximation (\ref{xmapprox}) and rewrite the result in terms of $\taucd$ using equation (\ref{taucdeqn}), we obtain
\begin{equation}
\label{mh2taucd}
\fmol = 1 - \frac{3\psi}{4 \taucd}.
\end{equation}
We must now express $\taucd$ in terms of $\psi$, $\taucomp$, and $a$ alone. Thus
\begin{eqnarray}
\taucd & = & \taucomp + a\psi\left(\frac{\taucomp}{\tauR}\right) \\
\label{taucdapprox}
& = & \taucomp + a\psi\left(\frac{\taucd}{\taurd}\right).
\end{eqnarray}
The second term on the RHS still involves the unknowns $\taucd$ and $\taurd$, but because they are already multiplied by $a$ we now need only determine them to zeroth order in $a$. Solviing equation (\ref{taucdeqn}) for $\taurd$ gives
\begin{eqnarray}
\frac{\taurd}{\taucd} & = & \frac{1}{\phimol}+\left(1-\frac{1}{\phimol}\right)\frac{3\psi}{4\taucd} \\
& \approx & \frac{1}{\phimol}+\left(1-\frac{1}{\phimol}\right)\frac{3\psi}{4\taucomp},
\label{taucdapprox1}
\end{eqnarray}
where in the second step we have dropped a term of order $a$ to obtain an expression that is accurate to zeroth order in $a$. Substituting this into equation (\ref{taucdapprox}), and thence into equation (\ref{mh2taucd}), gives our final expression for the molecular mass fraction, accurate to first order in $a$:
\begin{equation}
\label{molfracanalyt}
\fmol = 1 - \frac{3\psi}{4\taucomp} \left[1 + \frac{4a\psi\phimol}{4\taucomp+3(\phimol-1)\psi}\right]^{-1}.
\end{equation}
Comparison of this approximate expression with the numerical solution illustrated in Figure \ref{finitecloud} shows that for our fiducial $\phicnm=3$ and metallicities from $Z'=10^{-2}-10$, it is accurate to better than 30\% whenever the approximation analytic solution gives $\fmol > 0.25$, but that it goes to zero too sharply at low molecular fraction. We can improve the approximation by forcing the H$_2$ fraction to approach zero smoothly rather than sharply at low column density. Experimentation shows that the expression
\begin{equation}
\fatm^{-3} = 1+\left\{\left(\frac{4\taucomp}{3\psi}\right)\left[1 + \frac{4a\psi\phimol}{4\taucomp+3(\phimol-1)\psi}\right]\right\}^{3}
\label{hifracanalyt}
\end{equation}
matches the numerical result for $\fatm\equiv M_{\rm HI}/M$ for $\phicnm=3$ to better than 20\% for all $Z'<10$ regardless of the value of $\fatm$. (However note that, as we show in \S~\ref{geomerror}, for $\fmol\ltsim 1/2$ our estimate of the molecular content is only an upper limit, and this is true of equation \ref{hifracanalyt} as well.) Using equation (\ref{taucompdefn}) to replace $\taucomp$ with $\Sigmacomp$, and substituting in our fiducial values $\phicnm=5$, $a=0.2$, and $\sigmad=10^{-21}Z'$ cm$^{-2}$, equation (\ref{hifracanalyt}) becomes
\begin{equation}
\label{hifracanalyt1}
\fatm \rightarrow \left[1+\left(\frac{s}{11}\right)^3\left(\frac{125+s}{96+s}\right)^3\right]^{-1/3}
\end{equation}
where
\begin{equation}
s \equiv \frac{\Sigma_{\rm comp,0} Z'}{\psi}
\end{equation}
and $\Sigma_{\rm comp,0} = \Sigmacomp/(1\,\msun\mbox{ pc}^{-2})$. Note that our result indicates that to good approximation the molecular content of an atomic-molecular complex depends only on the combination of input parameters $Z'\Sigma_{\rm comp,0}/\psi$; the numerator $Z'\Sigma_{\rm comp,0}$ is simply the dust column density of the complex up to a scaling factor, while the denominator $\psi$ is the dimensionless radiation field, which equations (\ref{chieqn}) and (\ref{psidefn}) give solely as a function of metallicity.

From (\ref{hifracanalyt1}), it also immediately follows that the H$_2$ to \hi ratio $\rmol\equiv \fmol/\fatm$ is
\begin{equation}
\rmol \approx \left[1+\left(\frac{s}{11}\right)^3\left(\frac{125+s}{96+s}\right)^3\right]^{1/3}-1.
\end{equation}
For $\rmol>1$, which is the regime for which our models apply with high confidence, we can use an even simpler expression
\begin{equation}
\rmol \approx 0.08s = 0.08 \frac{\Sigma_{\rm comp,0} Z'}{\psi},
\end{equation}
where is accurate to $\sim 30\%$.

Similarly, we show in \citetalias{krumholz08c} that the dust absorption optical depth through the atomic layer for a finite cloud is well-approximated by
\begin{equation}
\tauhi \approx \frac{\psi}{4}\left[\frac{1}{1 - (a'/4)(\psi/\tauR)}\right],
\end{equation}
where $a'=\frac{3}{2}-4a=0.7$. If we treat $a'/4$ as a small parameter and perform a series expansion around it, then we need only approximate $\tauR$ to zeroth order in $a$. We can do this simply by using equation (\ref{taucdapprox1}) with $a=0$, which allows us to set $\tauR=\taurd$ and $\taucomp=\taucd$. Thus to zeroth order in $a$ we have
\begin{equation}
\tauR = \taurd \left[\frac{1}{\phimol}+\left(1-\frac{1}{\phimol}\right)\frac{3\psi}{4\taucomp}\right],
\end{equation}
and to first order in $a$ or $a'$ we have
\begin{equation}
\label{tauhianalyt}
\tauhi = \frac{\psi}{4}\left[1-\frac{a'\psi\phimol}{4\taucomp+3(\phimol-1)\psi}\right]^{-1}.
\end{equation}
As with approximation (\ref{molfracanalyt}) for the molecular mass fraction, this expression works well whenever the molecular fraction is not too low, and may be improved by forcing the optical depth to approach the total cloud optical depth smoothly when the column density becomes low. The expression
\begin{equation}
\tauhi^{-2} = \taucomp^{-2}+\frac{16}{\psi^2}\left[1-\frac{a'\psi\phimol}{4\taucomp+3(\phimol-1)\psi}\right]^{2}
\end{equation}
is accurate to better than 35\% for all $Z'<10$, and to better than 25\% for $Z'<1$.

It is also convenient to invert our analytic expressions to determine column density as a function of molecular content and metallicity. The term $(125+s)/(96+s)$ in equation (\ref{hifracanalyt1}) is generally close to unity except at extremely high column densities, so if we neglect second-order corrections to the difference between this term and unity, we can solve equation (\ref{hifracanalyt1}) for $\fatm$ to obtain
\begin{equation}
\label{sinveqn}
\frac{\Sigma_{\rm comp,0} Z'}{\psi} \rightarrow 11 \left(\fatm^{-3}-1\right)^{1/3} \frac{8.7+ \left(\fatm^{-3}-1\right)^{1/3}}{11+ \left(\fatm^{-3}-1\right)^{1/3}}
\end{equation}
This expression matches the numerical solution at the $\sim 20\%$ level for $\fhi<0.75$. Note that this result implies that the column density $\Sigmacomp$ at which a given \hi fraction is reached depends on metallicity both explicitly through the $Z'$ term in the numerator, representing the effect of metallicity on dust content, and implicitly through $\psi$ (equations \ref{chieqn} and \ref{psidefn}), representing the effect of metallicity on the ratio of radiation intensity to CNM density. Since $\psi$ is an increasing function of metallicity, for a given molecular fraction $\Sigmacomp$ has a weaker than linear dependence on metallicity. For example, evaluating (\ref{sinveqn}) with $\fhi=0.5$ at solar metallicity $Z'=1$ indicates that we expect the gas to be half molecular for complexes with $\Sigmacomp=27$ $\msun$ pc$^{-2}$. (The exact numerical solution is $\Sigmacomp=25.5$ $\msun$ pc$^{-2}$.) At one-third solar metallicity, $Z'=1/3$, half molecular content is reached at $\Sigmacomp=67$ $\msun$ pc$^{-2}$ (using equation \ref{sinveqn}; numerically $\Sigmacomp=55.3$ $\msun$ pc$^{-2}$), somewhat less than a factor of $3$ higher.

\section{Comparison to Observations}
\label{obscomparison}

Our model makes strong predictions for the relative fractions of \hi and H$_2$ as a function of total surface density and metallicity, and in this section we compare these results to a variety of galactic and extragalactic observations.

\subsection{Extragalactic Observations}

\subsubsection{Data Sets}

\begin{deluxetable}{lccc}
\tablecaption{Galaxy metallicities\label{metaltab}}
\tablewidth{0pt}
\tablehead{
\colhead{Galaxy} &
\colhead{$\log(\mbox{O}/\mbox{H})+12$\tablenotemark{a}} &
\colhead{Sample\tablenotemark{b}} &
\colhead{Reference}
}
\startdata
Solar (Milky Way) & 8.76 & B & 7 \\
DDO154	& 7.67 & L & 3 \\
HOI     & 7.54 & L & 6\\
HOII    & 7.68 & L & 6\\
IC10    & 8.26 & B & 1\\
IC2574  & 7.94 & L & 6\\
NGC0598 & 8.49 & B & 5 \\
NGC0628 & 8.51 & L & 5\\
NGC0925 & 8.32 & L & 5\\
NGC2403 & 8.39 & L & 5\\
NGC2841 & 8.81 & L & 5\\
NGC2976 & 8.30  & L & 8 \\
NGC3077 & 8.64 & L & 8 \\
NGC3184 & 8.72 & L & 5\\
NGC3198 & 8.42 & L & 5\\
NGC3351 & 8.80 & L & 5\\
NGC3521 & 8.49 & BL & 5\\
NGC3627 & 9.25 & BL & 4\\
NGC4214 & 8.22 & L & 2\\
NGC4321 & 8.71 & B & 5 \\
NGC4414 & ...  & B & ... \\
NGC4449 & 8.31 & L & 2\\
NGC4501 & 8.78 & B & 5 \\
NGC4736 & 8.50 & BL & 5\\
NGC5033 & 8.68 & B & 5\\
NGC5055 & 8.68 & BL & 5\\
NGC5194 & 8.75 & BL & 5\\
NGC5457 & 8.44 & B & 5\\
NGC6946 & 8.53 & L & 5\\
NGC7331 & 8.48 & BL & 5\\
NGC7793 & 8.34 & L & 5\\
\enddata
\tablenotetext{a}{We take the metallicity relative to solar to be proportional to the $\mbox{O}/\mbox{H}$ ratio, i.e.\ $\log Z' = [\log(\mbox{O}/\mbox{H})+12]-8.76$.}
\tablenotetext{b}{B = galaxy is in \citetalias{blitz06b} sample, L = galaxy is in \citetalias{leroy08} sample}
\tablerefs{
1 -- \citet{garnett90}, 2 -- \citet{martin97}, 3 -- \citet{vanzee97}, 4 -- \citet{ferrarese00}, 5 -- \citet{pilyugin04}, 6 -- \citet{walter07}, 7 -- \citet{caffau08}, 8 -- \citet{walter08a}. An entry ... indicates that there is no gas-phase metallicity is reported in the literature.
}
\end{deluxetable}

We use three extragalactic data sets for comparison to our models. Two are recent surveys that have mapped nearby galaxies in 21 cm \hi and 2.6 mm CO($1\rightarrow0$) emission at overlapping positions, and therefore provide an ideal laboratory in which to test our model. The first of these is the work of \citetalias{wong02}, \citetalias{blitz04}, and \citetalias{blitz06b}, who report \hi and H$_2$ surface densities on a pixel-by-pixel basis in 14 nearby galaxies (including the Milky Way). The H$_2$ surface densities are inferred from CO observations taken as part of the BIMA SONG survey \citep{regan01, helfer03}, while the \hi observations are from the VLA. \citetalias{wong02} and \citetalias{blitz06b} supplement these data with stellar surface density measurements from 2MASS \citep{jarrett03}, which together with the equations given in \citetalias{blitz06b} can be used to derive a pressure in each pixel if one assumes that the gas in in hydrostatic balance, that the stellar scale height greatly exceeds the gas scale height, and that the gas velocity dispersion has a known value. The galaxies in the sample are all molecule-rich spirals with metallicities within $0.5$ dex of solar.

The second extragalactic data set we use is compiled by \citetalias{leroy08}, who a combine \hi measurements from the THINGS survey with CO data partly taken from BIMA SONG and partly from the ongoing HERACLES survey. The authors also include 2MASS stellar surface densities in their compilation, and give mean pressures. Unlike the \citetalias{blitz06b} sample the data reported are averages over galactocentric rings rather than individual pixels, although point-by-point maps at sub-kpc resolution are in preparation (F.~Walter, 2008, private communication). The sample includes 23 galaxies, of which roughly half are large spirals and roughly half are low-mass, H\textsc{i}-dominated dwarfs. The galaxies in the data set partly overlap with those in the sample of \citetalias{blitz06b}, but extend over a wider range of metallicities and molecule fractions.

We summarize the galaxies in the samples in Table \ref{metaltab}. We also report metallicities for each galaxy where these are available in the literature. We do not include NGC4414 in the analysis, because no gas phase metallicity is available for it. In the comparison that follows, we neglect the presence of metallicity gradients within these galaxies, because gradients are only available for some of them. On top of this, we note that the metallicities themselves are probably uncertain at levels from hundreths to tenths of a dex, depending on the galaxy and the analysis technique used, which adds additional scatter on top of that already introduced by our neglect of metallicity gradients.

In both data sets the uncertainty in the \hi column densities is generally $\sim 10\%$. In the CO data the formal uncertainties are generally $\sim 20-30\%$, but the dominant uncertainty is probably systematic: the X factor used to convert observed CO luminosities into H$_2$ column densities. This is uncertain at the factor of $\sim 2$ level \citep{blitz07a}, and almost certainly varies with metallicity \citep[e.g.][]{bolatto08a}. Thus, although for clarity we will suppress error bars in the plots that follow, recall that the molecular data is uncertain at the factor of $\sim 2$ level.

The third data set we use, in \S~\ref{smcsec}, is the {\it Spitzer} Survey of the Small Magellanic Cloud (S$^3$MC) \citep{bolatto07a, leroy07a}. This survey differs from the SONG and THINGS data sets in that those surveys infer the presence of H$_2$ via CO emission, whereas S$^3$MC measures molecular hydrogen using measurements of dust from {\it Spitzer} combined with \hi measurements by \citet{stanimirovic99a} and \citet{stanimirovic04a}. The basic idea behind the technique is that one determines the dust-to-gas ratio in a low-column density region where molecules are thought to contribute negligibly to the total column. Then by comparing the \hi and dust column density maps, one can infer the presence of H$_2$ in pixels where the total dust column exceeds what one would expect for the observed \hi column and a fixed dust-to-gas ratio. The reason for using this technique is that, at the low metallicity of the SMC ($\log[\mbox{O}/\mbox{H}]+12=8.0$, \citealt{dufour84a}, i.e.\ $0.76$ dex below Solar), CO may cease to be a reliable tracer of molecular gas. The SMC represents the lowest metallicity galaxy for which we have H$_2$ detections rather than upper limits; the \citetalias{blitz06b} sample does not contain any galaxies with metallicities as low as the SMC, and no CO was detected in any of the \citetalias{leroy08} galaxies with metallicities comparable to or lower than the SMC. Thus, the SMC represents a unique opportunity to test our models at very low metallicity.

Before comparing to these data sets, it is worth commenting briefly on one additional extragalactic data set to which we will not compare our models: observations of H$_2$ column densities along sightlines in the LMC and SMC using {\it FUSE} \citep{tumlinson02a}. We do not use this data set for comparison because it includes only sightlines with low column densities that are strongly dominated by atomic gas; the highest reported H$_2$ fraction is below 10\%. Sightlines with significantly higher molecular content than this absorb too much background starlight to allow {\it FUSE} to make a reliable measurement of the H$_2$ column. The low column densities of the clouds that {\it FUSE} can observe in these galaxies place almost all of them into the regime where our theory yields only upper limits. Those limits are generally consistent with the data, but the comparison is not particularly illuminating.

\subsubsection{Column Density and Metallicity Dependence}
\label{sigmazdepend}

In this section we compare our models predictions of the \hi and H$_2$ mass fractions as a function of total column density and metallicity to our two extragalactic data sets. To perform this comparison, for convenience we first break the sample into four metallicity bins: $-1.25 < \log Z' < -0.5$, $-0.5 < \log Z' < -0.25$, $-0.25 < \log Z' < 0$, and $0 < \log Z' < 0.5$, where $Z' = Z / Z_{\odot}$ and we adopt $\log(\mbox{O}/\mbox{H})+12 = 8.76$ as the value corresponding to solar metallicity \citep{caffau08}. These bins roughly evenly divide the data; we do not include the galaxies for which metallicities are not available in the literature.

Next we must consider how finite spatial resolution will affect our comparison. Given the beam sizes in the observed data, which range from $\sim 0.3$ kpc for the nearest galaxies to a few kpc for the most distant,  each pixel (for \citetalias{blitz06b}) or ring (for \citetalias{leroy08}) in the observed data set is likely to contain multiple atomic-molecular complexes. The atomic and molecular observations are convolved to the same resolution, so this does not bias the measurement of the atomic to molecular ratio, although it does mean that this ratio is measured over an averaging scale set by the beam size. Since individual atomic molecular complexes presumably represent peaks of the galactic column density, however, the total (\hi + H$_2$) observed gas surface density $\Sigmaobs$ reported for each pixel or ring represents only a lower limit on $\Sigmacomp$, the total surface density of individual complexes. In our model the fraction of the cloud in molecular form is a strictly increasing function of $\Sigmacomp$ (and of metallicity $Z$), while the atomic fraction is a strictly decreasing function. Because we have only the observed column density $\Sigmaobs$ available, not the column density of an individual complex $\Sigmacomp$, we have no choice but to take $\Sigmacomp = \Sigmaobs$ as our best estimate. The error in this approximation probably ranges from tens of percent for nearby galaxies where the beam size is not much larger than the size of a complex up to an order of magnitude or more for ring-averages in distant galaxies. As a result, though, we expect to overestimate the atomic fraction and underestimate the molecular fraction. Physically we may think of this as a clumping effect: the clumpier the gas is, the better able it is to shield itself against dissociating radiation. Since our observations smooth over scales larger than the characteristic gas clumping scale, we will miss this effect.

There are also two effects which go in the other direction, however. First, as noted above, shielding comes primarily from cold atomic gas, not warm gas. Although most of the gas in the immediate vicinity of a single molecular region is probably cold, observations that average over many molecular cloud complexes are likely to include a fair amount of WNM gas as well. Since we have only the total \hi column densities including both phases, we are prone to overestimate $\Sigmacomp$ and therefore the molecular fraction, because the WNM raises the \hi column but does not provide much shielding. This effect is not important at moderate to high column densities, where the atomic gas does not dominate the total mass budget, but it could be significant at lower densities.

Second, there may be gas along our line of sight through a galaxy that is not associated with an atomic-molecular complex along that line of sight. This effect will increase in severity as the galaxy comes close to edge-on, since this will increase the path length of our line of sight through the galaxy. As with WNM gas, this extra material contributes to $\Sigmaobs$ but not to the complex surface density $\Sigmacomp$, and it therefore leads us to overestimate the molecular fraction. We can perform a very simple calculation to estimate the size of this effect. Consider a simple self-gravitating gas disk characterized by the standard vertical density profile $n \propto \sech^2(0.88 z/h)$, where $h$ is the half-height of the gas. At some point in this disk is an atomic-molecular complex, centered at the midplane. Since complexes are found preferentially at the midplane, unless the galaxy is very close to edge-on then we need not consider the possibility of our line of sight intersecting multiple independent complexes. Since the complex formed from a large-scale gravitational instability in the disk, its characteristic size is $\sim h$, and we therefore consider gas to be ``associated" with the complex if it is within a distance $h$ of it in the plane of the disk. Suppose this galaxy has an inclination $i$. Making the worst-case assumption, that there is no density enhancement due to the presence of the self-gravitating complex, we can then compute what the fraction of the gas we see along our line of sight is not associated with it. This is simply
\begin{equation}
\frac{\int_{h\cot i}^{\infty} \sech^2 (0.88 z/h) \, dz}{\int_0^{\infty} \sech^2 (0.88 z/h) \,dz}
= 1 - \tanh(1.13 \cot i).
\end{equation}
This is less than $0.5$ for all inclinations less than $64^{\circ}$. Only a handful of the galaxies in the SONG and THINGS surveys have inclinations larger than this, so even in the worst case scenario where complexes do not represent any enhancement of the gas density, we expect non-associated gas to produce an error smaller than a factor of $\sim 2$ for the great majority of the galaxies to which we are comparing.

Given the limitations imposed by finite resolution, we proceed as follows. Using the method described in \S~\ref{finite}, we compute the molecular mass fraction
\begin{equation}
f_{\rm H_2}(\Sigmacomp, Z) \equiv \frac{M_{\rm H_2}}{M},
\end{equation}
as a function of complex column density $\Sigmacomp$ and metallicity $Z$. (For these and all subsequent predictions we use our fiducial value of $\phicnm=3$.) We then generate predicted \hi and H$_2$ column densities for each metallicity bin and each observed column density $\Sigma$ via
\begin{eqnarray}
\label{h2mfraceqn}
\Sigma_{\rm H_2,predicted} & = & f_{\rm H_2}(\Sigmaobs, Z_{\rm min}) \Sigmaobs \\
\Sigma_{\rm HI,predicted} & = & [1-f_{\rm H_2}(\Sigmaobs, Z_{\rm min})] \Sigmaobs,
\end{eqnarray}
where $Z_{\rm min}$ is the minimum metallicity for that bin. Since $f_{\rm H_2}(\Sigmacomp, Z)$ is a strictly increasing function of $\Sigmacomp$ and $Z$, and we know that $\Sigmaobs < \Sigmacomp$ and $Z_{\rm min} < Z$, we expect $f_{\rm H_2}(\Sigmaobs, Z_{\rm min})$ to be a lower limit on the true molecular fraction. We therefore expect $\Sigma_{\rm H_2,predicted}$ to be a lower limit on the observed data, and $\Sigma_{\rm HI,predicted}$ to be an upper limit. The possible exception to this statement is at low column densities, where a significant fraction of the \hi column may be in the form of WNM.

\begin{figure}
\plotone{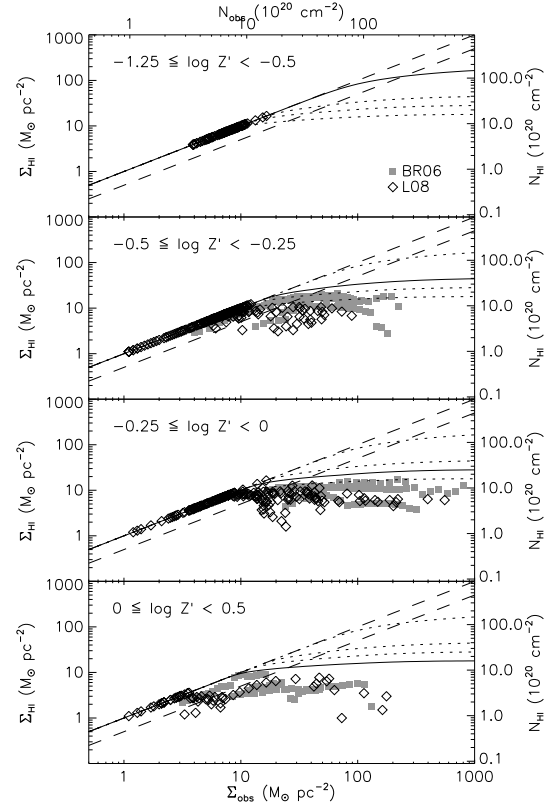}
\caption{
\label{hicomp}
\hi column density $\Sigma_{\rm \hi}$ versus total column density $\Sigma$ for galaxies in metallicity bins $-1.25 < \log Z' < -0.5$, $-0.5 < \log Z' < -0.25$, $-0.25 < \log Z' < 0$, and $0 < \log Z' < 0.5$, as indicated. In each panel we plot the values of $\Sigma_{\rm \hi}$ and $\Sigma$ from the samples of \citetalias{blitz06b} (\textit{squares}) and \citetalias{leroy08} (\textit{diamonds}) and show our model predictions of $\Sigma_{\rm \hi}$ as a function of $\Sigma$ for $\log Z' = -1.25$, $-0.5$, $-0.25$, and $0$ (\textit{lines, highest to lowest}). The curve for the value of $\log Z'$ equal to the minimum $\log Z'$ for each bin, which should represent the upper envelope of the data, is shown as a solid line. The rest are shown as dotted lines. The four lines are the same in each panel. To maximize readability we omit the error bars on the data points. The parallel slanted dashed lines, $\Sigmahi=\Sigmaobs$/2 and $\Sigmahi = \Sigmaobs$, show the range of value of $\Sigmahi$ for which our predictions should be treated as upper limits for the reasons discussed in \S~\ref{geomerror}. Parts of our model curves above the $\Sigmahi=\Sigmaobs/2$ line could be as high as the $\Sigmahi = \Sigmaobs$ line.
}
\end{figure}

\begin{figure}
\plotone{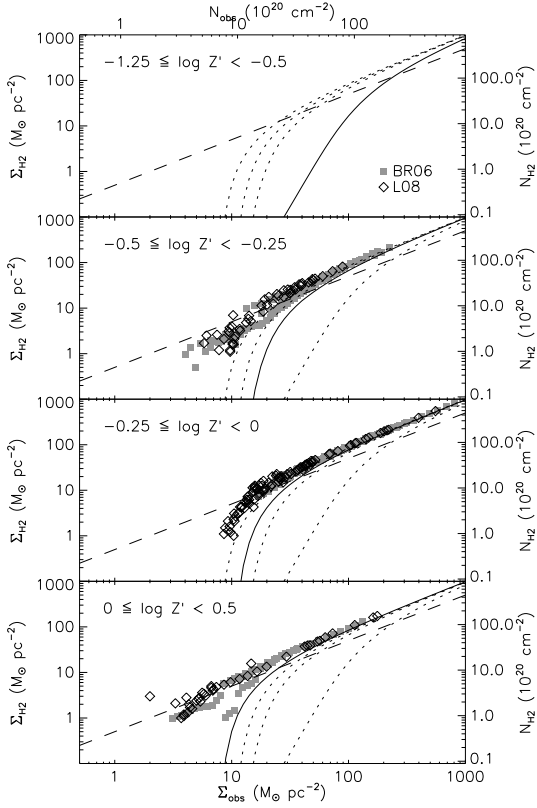}
\caption{
\label{molcomp}
Same as Figure \ref{hicomp}, except that we show H$_2$ rather than \hi column densities. For clarity we do not show non-detections. Note that $N_{\rm H_2}$ is the column density of H nuclei in molecular form, which is twice the column density of H$_2$ molecules. The slanted dashed line is $\Sigma_{\rm H_2}=\Sigmaobs/2$; as discussed in \S~\ref{geomerror}, above this line our model curves may be taken as predictions, while below it they should be taken as upper limits.
}
\end{figure}

\begin{figure}
\plotone{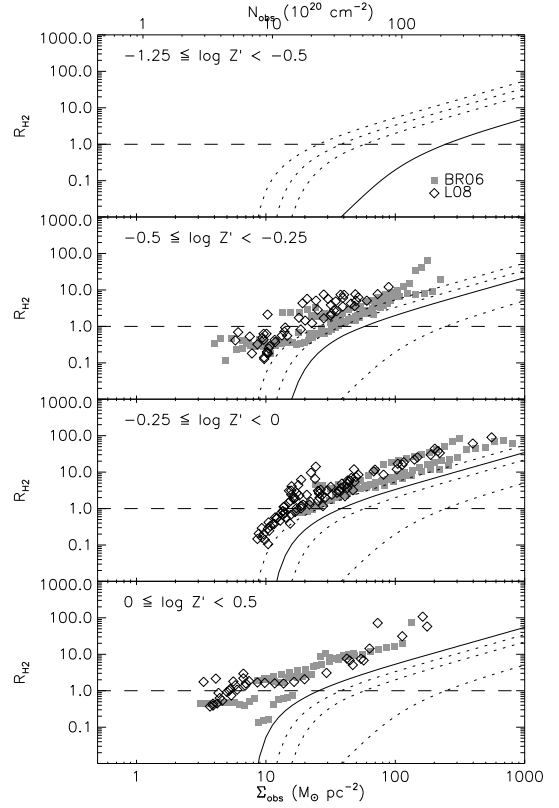}
\caption{
\label{rmolcomp}
Same as Figures \ref{hicomp} and \ref{molcomp}, except that we show $R_{\rm H_2}\equiv \Sigma_{\rm H_2}/\Sigmahi$ rather than column densities. For clarity we do not show non-detections. The horizontal dashed line corresponds to $R_{\rm H_2} = 1$; as discussed in \S~\ref{geomerror}, above this line our model curves may be taken as predictions, while below it they should be taken as upper limits.
}
\end{figure}

We plot the data against our theoretical prediction for the upper envelope of $\Sigmahi$ in Figure \ref{hicomp}, and we show the corresponding predicted lower envelopes for $\Sigma_{\rm H_2}$ and $R_{\rm H_2}\equiv \Sigma_{\rm H_2}/\Sigmahi$ in Figures \ref{molcomp} and \ref{rmolcomp}. For the \citetalias{blitz06b} data set, rather than plotting the tens of thousands of individual pixels it contains, for each galaxy we show the data averaged over 20 logarithmically-spaced column density bins running from the minimum to the maximum value of $\Sigmaobs$ reported for that galaxy. As the plots show, our model predictions for the upper envelope of the \hi surface density and the corresponding lower envelope of the H$_2$ surface density as a function of total surface density and metallicity agree very well with the data. The data fill the space up to our predicted envelopes but for the most part do not cross them, even when the predicted \hi envelope becomes flat for $\Sigmaobs \gtsim 10$ $\msun$ pc$^{-2}$. Moreover, our model recovers not only the primary dependence of $\Sigmahi$ on the total observed column density $\Sigmaobs$, but also the secondary dependence on metallicity. For example, rings in the lowest metallicity bin in the \citetalias{leroy08} data set reach total mean column densities of almost $20$ $\msun$ pc$^{-2}$, but still show no detectable molecular component. On the other hand, in the highest metallicity bin the molecular fraction is close to $70\%$ in rings with $\Sigmaobs\approx 20$ $\msun$ pc$^{-2}$. Our models reproduce this effect: at a metallicity of $\log Z'=-0.5$, we predict that the gas will be $94\%$ atomic even at a surface density of $20$ $\msun$ pc$^{-2}$, whereas for $\log Z'=0.5$ we predict an atomic fraction of only $30\%$ at that column density, in agreement with the data.

We caution that we can only predict upper limits on the molecular content in regions of parameter space when our predicted molecular fraction falls below $\sim 1/2$, for the reasons discussed in \S~\ref{geomerror}. We have indicated the regions where our model predictions convert to upper limits in Figures \ref{hicomp} -- \ref{rmolcomp}. Alternately, one can express this uncertainty as giving a minimum column density at which we can predict a value rather than an upper limit for molecular content. For reference, at the metallicities of $\log Z'=-1.25$, $-0.5$, $-0.25$, $0$, and $0.5$ which define the edges of our metallicity bins, the minimum column densities for which we can predict numerical values to better than factor of few confidence are $250$, $58$, $38$, $25$, and $11$ $\msun$ pc$^{-2}$, respectively.

\begin{figure}
\plotone{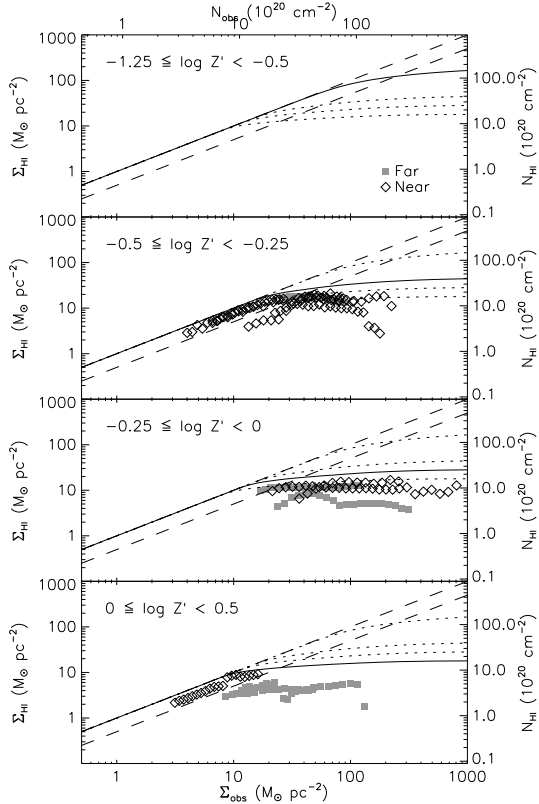}
\caption{
\label{hicomp_dist}
Same as Figure \ref{hicomp}, except that we show only data from \citetalias{blitz06b}, and we have divided this data set into ``near" galaxies, those for which the survey resolution is $<1$ kpc, and ``far" galaxies, those for which the resolution is $>1$ kpc.
}
\end{figure}

Finally, we note that the future HERACLES / THINGS data set represents an opportunity to perform an even stronger test of our model. In Figure \ref{hicomp} a significant fraction of the data points fall below our predicted upper limits, and correspondingly these points are above our lower limits in Figures \ref{molcomp} and \ref{rmolcomp}. We hypothesize that these data points represent rings or pixels within which the gas is significantly clumped, so that the averaged column density $\Sigmaobs$ seen in the observation is significantly lower than the column density at which most of the molecular gas in that beam or ring is found, and our calculation for a complex with $\Sigmacomp=\Sigmaobs$ overestimates the \hi column. In reality these regions probably consist of patches of high column density where most of the molecules reside, embedded in a lower density ambient medium that has a lower molecular fraction than we determine by averaging over large scales. If we could observe these regions at higher resolution, in Figure \ref{hicomp} the high column, high molecular content points will lie to the right of and slightly above the low resolution points, since both the total and \hi column densities will be higher than for the lower resolution observation, but the increase in the total column will be larger than in the \hi column. Conversely, the low column, low molecule patches will lie to the left and slightly downward from the low resolution points, since both the total and \hi columns will decline, but the \hi by less, since the atomic fraction rises. These changes will bring the data points closer to our model curves. Indeed, the existing data already hint that such an effect is present: the single beam-averaged observations of the \citetalias{blitz06b} data set scatter away from our limit lines noticeably less than the ring-averaged observations from \citetalias{leroy08}. Similarly, we can divide the point-by-point data from \citetalias{blitz06b} into a ``near" sample, consisting of galaxies for which the resolution is smaller than 1 kpc, and a ``far" sample, consisting of galaxies with larger resolutions. We plot a version of Figure \ref{hicomp} using only this divided data set from \citetalias{blitz06b} in Figure \ref{hicomp_dist}. The comparison is quite noisy, but the data in that bin do at least seem consistent with the hypothesis that the near data fall closer to the model lines than the far data.

The full HERACLES survey, currently underway, will report measurements of the molecular surface density for individual patches $\sim 0.5$ kpc in size, generally smaller than the beam patches the BIMA / SONG survey. While this is still considerably larger than the $\ltsim 0.1$ kpc-size of a typical atomic-molecular complex in the Milky Way, at the higher spatial resolution of the full HERACLES data set beam-smearing effects should be reduced and the column densities reported for each patch should be closer to the true column densities $\Sigmacomp$ of the individual atomic-molecular complexes. We therefore predict that in the full HERACLES data set the $\Sigma_{\rm \hi} - \Sigmaobs$ relation should be closer to our theoretical upper and lower limit curves than the lower resolution or azimuthally-averaged data shown in Figures \ref{hicomp} -- \ref{rmolcomp}. 

\subsubsection{Pressure Dependence}

\citetalias{blitz04} and \citetalias{blitz06b} find that the molecular fraction in a galaxy correlates with the mid-plane gas pressure as
\begin{equation}
\rmol \equiv \frac{\Sigma_{\rm H_2}}{\Sigmahi} = \left(\frac{P/\kb}{3.5\pm0.6\times 10^4\mbox{ K cm}^{-3}}\right)^{0.92\pm 0.07},
\label{br06fit}
\end{equation}
while \citetalias{leroy08} find
\begin{equation}
\rmol = \left(\frac{P/\kb}{2.0\times 10^4\mbox{ K cm}^{-3}}\right)^{0.8}
\label{l08fit}
\end{equation}
for their sample. Although our model does not directly give a prediction for the dependence of the molecular fraction on pressure, and we argue that surface density and metallicity are the physical variables that directly control the molecular fraction, we wish to check whether our model is consistent with the observed correlation. 

To avoid introducing any bias in performing this check, we must determine pressures from observable quantities in the same way that \citetalias{blitz06b} and \citetalias{leroy08} do, following an approximation introduced by \citetalias{blitz04}.\footnote{It is important to note that the midplane pressure which the observers attempt to estimate is not the same as the pressure in the molecular gas or the CNM which we balanced in \S~\ref{normrad}; the total midplane pressure includes contributions from all phases of the ISM rather than just the CNM or molecular clouds, as well as contributions from magnetic fields, cosmic rays, and bulk motions. The pressure we use in \S~\ref{normrad} includes thermal pressure in the CNM or molecular gas only.} The BR04 approximation treats the galaxy as an infinite thin disk of uniform gas and stars in vertical hydrostatic balance. For such a disk the pressure is related to surface density by 
\begin{equation}
\label{psigma}
P = \frac{\pi}{2} G \Sigmag \left(\Sigmag + \Sigma_*\frac{\vg}{v_*}\right),
\end{equation}
where $\Sigmag$ is the total (\hi plus H$_2$) gas surface density, $\Sigma_*$ is the stellar surface density, and $\vg$ and $v_*$ are the vertical velocity dispersions of the gas and stellar components, respectively. The gas velocity dispersion is roughly constant, $\vg \approx 8$ km s$^{-1}$, for the galaxies in the \citetalias{blitz06b} and \citetalias{leroy08} samples. The term $ \Sigma_*/v_*$ presents more difficulty, however. It varies by orders of magnitude from galaxy centers to edges, and between galaxies, so we cannot pick a single value for it. Since in our model the molecular fraction is a function only of metallicity and $\Sigmag$, this means that we do not predict a single-valued relationship between $\rmol$ and $P$ that we can directly compare to the empirical fits.

However, we can still compare our model to the data in two ways. First, we can pick a range of values of $\Sigma_*/v_*$ consistent with the range in the observed sample, and demonstrate that the resulting range of predictions for $\rmol$ covered by our model is consistent with the observational data. For this purpose we follow \citetalias{blitz06b} in adopting a constant value $\vg=8$ km s$^{-1}$ and using simplified version of equation (\ref{psigma}), which follows from assuming that both the surface density and scale height of the stars are much greater than those of the gas:
\begin{eqnarray}
\frac{P}{\kb} & = & 272\mbox{ cm}^{-3}\mbox{ K} \left(\frac{\Sigmaobs}{\msun\mbox{ pc}^{-2}}\right)
\left(\frac{\Sigma_*}{\msun\mbox{ pc}^{-2}}\right)^{0.5} 
\nonumber
\\
& & \qquad {} \times \left(\frac{\vg}{\mbox{km s}^{-1}}\right) \left(\frac{h_*}{\mbox{pc}}\right)^{-0.5},
\label{presdefn}
\end{eqnarray}
where $\Sigmaobs$ is the observed value of the total gas surface density $\Sigmag$, $h_*$ is the stellar scale height, and for convenience we define $\rho_*=\Sigma_*/h_*$. We consider values of $\log \rho_*$ from $-2.5$ to $1.5$ in units of $\msun$ pc$^{-3}$, a range that covers almost all of the samples of \citetalias{blitz06b} and \citetalias{leroy08}. For a given value of $\rho_*$, equation (\ref{presdefn}) gives a single-valued relationship between $P$ and $\Sigmag$, so for each $P$ and a choice of metallicity we can use our model to make a prediction for $\Sigma_{\rm HI}$ and thus for $\rmol$. As with our calculations in \S~\ref{sigmazdepend}, we expect to systematically underpredict $\rmol$ because the observed column density $\Sigmaobs$ is an underestimate of the true column density of atomic-molecular complexes that are not resolved by the telescope beam. 

\begin{figure}
\plotone{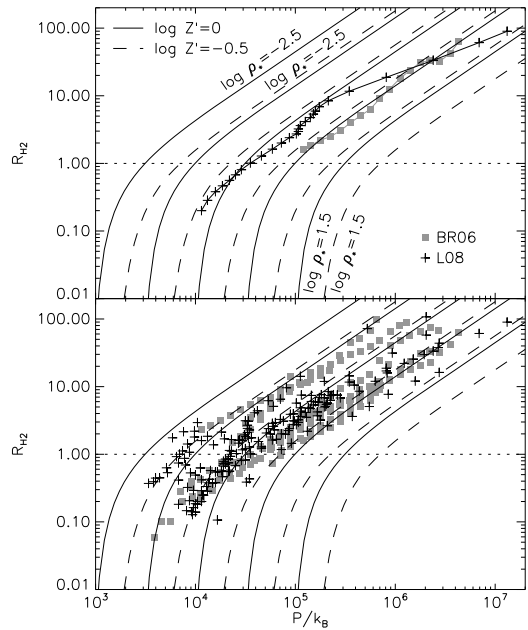}
\caption{
\label{pres1}
Molecular to atomic ratio $\rmol$ versus pressure $P$. We show the data sets of \citetalias{blitz06b} (\textit{squares}) and \citetalias{leroy08} (\textit{diamonds}), and predictions for models with $\log \rho_*=-2.5$, $-1.5$, $-0.5$, $0.5$, and $1.5$ in units of $\msun$ pc$^{-3}$ (\textit{highest to lowest lines, as indicated}), and with metallicities $\log Z'=-0.5$ and $0$ (\textit{dashed and solid lines, respectively}). In the upper panel for clarity we show only a single galaxy from each data set (NGC4736 from \citetalias{blitz06b} and HOII from \citetalias{leroy08}), while in the lower panel we show the data for all galaxies; the model curves in each panel are identical. For \citetalias{blitz06b} the data we show are averaged over 0.1 dex-wide bins in pressure, as in \citetalias{blitz06b}'s Figure 3. For the \citetalias{leroy08} data, we only plot rings for which both molecular and atomic gas are detected. The dotted horizontal lines show $\rmol=1$; below this line our predictions should be taken as upper limits only.
}
\end{figure}

We show the results of this computation in Figure \ref{pres1}. As expected, the model covers an area consistent with the observed data. First note that there is a systematic offset between the results from \citetalias{blitz06b} and \citetalias{leroy08}; this is likely because some of the structural parameters that are used in equation (\ref{presdefn}) are quite uncertain, particularly $h_*$, and the values adopted by the two surveys are not necessarily the same. We should keep this uncertainty in mind as we proceed, since it suggests an upper limit on the level of agreement we can expect.

Nonetheless, the correlation between $\rmol$ and $P$ shown in both observational surveys, $\rmol\propto P^{0.8-0.9}$, is somewhat flatter than any of our model curves for a particular choice of $\rho_*$, which all approach $\rmol \propto P^1$ at large $P$. However this is to be expected: low pressures and gas column densities are found preferentially in the outer parts of galaxies where $\rho_*$ is small, so at low $P$ we expect to be closer to the higher model curves, which have low $\rho_*$. Conversely, high pressures and values of $\Sigmag$ are found preferentially in regions with higher stellar densities $\rho_{*}$, so high $P$ values should be closer to the lower model curves, which have larger $\rho_*$. This covariance between $P$ and $\rho_*$ results in a flattening of the $\rmol \propto P$ relation that we would predict if all galaxies had fixed $\rho_*$.

We can eliminate this covariance effect and produce a stronger test than that shown in Figure \ref{pres1} by using our model to generate predictions for molecular content directly, and plotting those against the inferred pressure. To do this, for each galactocentric ring in the \citetalias{leroy08} data set we take the observed total gas surface density $\Sigmaobs$ and the metallicity given in Table \ref{metaltab} (which we treat as constant in each the galaxy) and use our model to predict $\rmol$. Similarly, for the \citetalias{blitz06b} data set we take the observed gas surface density $\Sigmaobs$ in each pixel and use our model to generate a prediction for $\rmol$. Since the pressure in each ring or pixel is known, by this procedure we generate a synthetic set of data points $(P,\rmol)$ which we can compare to the observations, and from which we can generate a fit for $\rmol$ as a function of $P$.

Before making the comparison, however, we must also model the effects of finite telescope sensitivity, which make it impossible to detect molecules below a certain minimum column density.  Since, as Figure \ref{pres1} shows, our model predicts that the slope of $\rmol$ versus $P$ varies with $P$, this can affect the fit we generate from our synthetic data. We include this effect using a procedure nearly identical to that of \citetalias{blitz06b}. These authors first estimate the minimum value of $\rmol$ as a function of pressure that they could detect in each galaxy based on their telescope noise limits. For our synthetic data we adopt a minimum $R_{\rm H_2,min}=0.05$ for all galaxies, slightly below the lowest average $\rmol$ that \citetalias{blitz06b} report for any of the galaxies they analyze. For each galaxy we then identify the lowest pressure $P_{\rm min}$ for which all the pixels / rings have $\rmol>R_{\rm H_2,min}$. At pressures above $P_{\rm min}$, the sample should be nearly complete, in the sense that no pixels or rings will have molecular non-detections that could bias a fit of $\rmol$ versus $P$. At pressures below $P_{\rm min}$ the data are incomplete, and we therefore drop all pixels or rings whose pressures are below $P_{\rm min}$. This produces a sensitivity-corrected set of synthetic data. Finally, following \citetalias{blitz06b} we then bin the pixels by pressure, using bins 0.1 dex wide, and for each pressure bin in each galaxy we compute an average value of $\rmol$. We do not bin the rings in the \citetalias{leroy08} sample.

\begin{figure}
\plotone{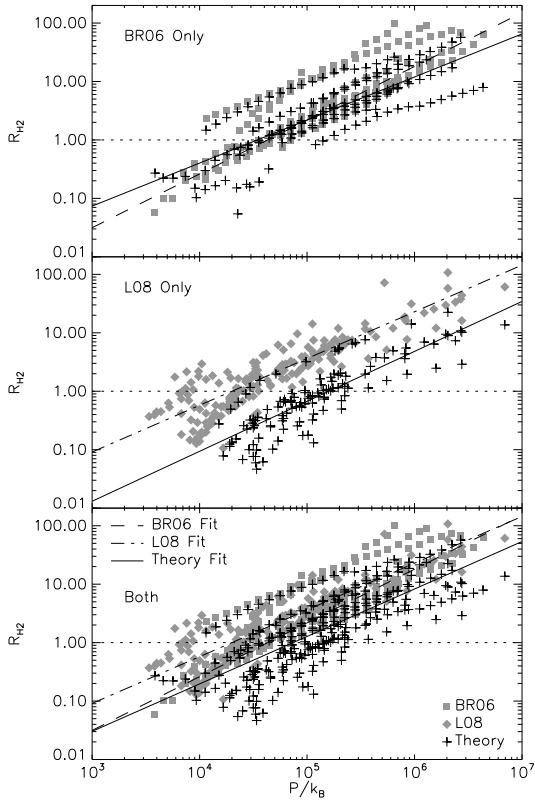}
\caption{
\label{pres2}
Molecular to atomic ratio $\rmol$ versus pressure $P$. We show the data sets of \citetalias{blitz06b} (\textit{squares}) and \citetalias{leroy08} (\textit{diamonds}), and predictions from theory for the value of $\Sigmag$ and $Z'$ for each data point (\textit{plus signs}). In the top panel we show only the \citetalias{blitz06b} data and the theoretical predictions corresponding to it. In the middle we show only the \citetalias{leroy08} data and predictions, and in the bottom panel we show both data sets together.
In all panels we only plot rings or points for which both molecular and atomic gas are detected for the observations, and the theoretical points are corrected for finite sensitivity as described in the text. We do not plot theoretical predictions for galaxies without measured metallicities. We also show the best powerlaw fits of \citetalias{blitz06b} data (equation \ref{br06fit}; \textit{dashed line}), the \citetalias{leroy08} data (equation \ref{l08fit}; \textit{dot-dashed line}), and to the model predictions (equation \ref{theoryfit}; \textit{solid line}). In the upper panel the theory line uses only the \citetalias{blitz06b} data, in the middle it uses only the \citetalias{leroy08} data, and in the bottom panel we show a line fit to our predictions for both data sets. The dotted horizontal lines show $\rmol=1$; below this line our predictions should be taken as upper limits only.
}
\end{figure}

We overplot the synthetic and real data in Figure \ref{pres2}. Fitting our synthetic data to a powerlaw function for $\rmol$ versus $P$ gives
\begin{equation}
\rmol = \left(\frac{P/\kb}{9.0\times 10^3\mbox{ K cm}^{-3}}\right)^{0.81},
\label{theoryfit}
\end{equation}
if we include our models for both the \citetalias{blitz06b} and \citetalias{leroy08} data sets. Using only one or the other gives
\begin{equation}
\rmol = 
\left\{
\begin{array}{ll}
\left(\frac{P/\kb}{2.2\times 10^2\mbox{\small\ K cm}^{-3}}\right)^{0.74}, \quad & \mbox{(BR06)} \\
\left(\frac{P/\kb}{2.8\times 10^5\mbox{\small\ K cm}^{-3}}\right)^{0.85}, \quad & \mbox{(L08)}
\end{array}
\right..
\end{equation}
As expected, we obtain a slope shallower than unity as a result of the systematic increase of $\rho_*$ with $P$. The scatter in the real and synthetic data are also comparable. As the Figure shows, our model predictions overlap with the observed data reasonably well, particularly for the \citetalias{blitz06b} data; the best fit for the \citetalias{blitz06b} data set is reasonably close to the observational best fit, while for \citetalias{leroy08} the theoretical predictions and the best fit to them give a slope similar to the observed value, but are systematically shifted to lower $\rmol$. The displacement between our model and the \citetalias{leroy08} data probably exists for the same reason that our predictions for $\Sigma_{\rm H_2}$ and $\rmol$ in \S~\ref{sigmazdepend} are lower limits: spatial averaging leads to an underestimate of the true column densities of the atomic-molecular complexes in a galaxy, which in turn leads us to slightly underpredict the molecular content. The averaging is significantly worse if done over rings than over individual beam pointings, so the offset is noticeably larger for the ring-averaged data set.

It is also worth cautioning that in performing this fit we have included model points where we predict $\rmol<1$, a region of parameter space where the predictions of our theory should be taken as upper limits. If we exclude these points from both our model predictions and from the observed data, all the slopes become significantly more shallow, but they remain consistent with one another. In this case our model predictions give a best fit of $\rmol = [(P/\kb)/(83\mbox{ K cm}^{-3})]^{0.47}$ for the combined BR06 and L08 samples, while the \citetalias{blitz06b} and \citetalias{leroy08} data sets give $[(P/\kb)/(120\mbox{ K cm}^{-3})]^{0.51}$ and $[(P/\kb)/(300\mbox{ K cm}^{-3})]^{0.53}$, respectively. These fits are performed without weighting by the errors or properly including the effects of upper limits, so they should only be taken as general indications, but the results do show that even if we limit our fit to the part of parameter space where we can apply our theory with high confidence, we obtain good agreement between model predictions and observations.

\begin{figure*}
\plotone{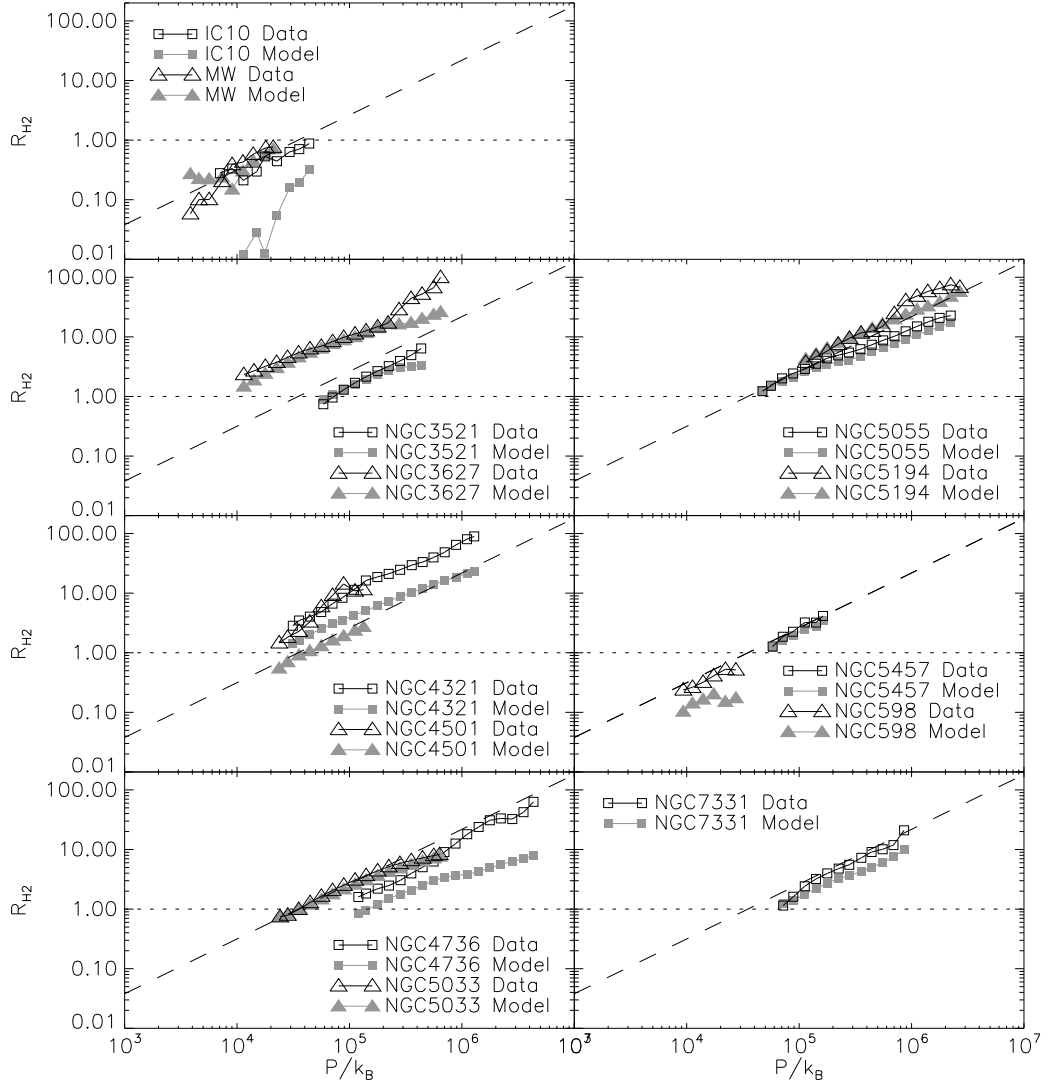}
\caption{
\label{pres_single}
Same as Figure \ref{pres2}, except that we show only the \citetalias{blitz06b}, and we plot only two galaxies per panel, as indicated, so that we can compare on a galaxy-by-galaxy basis. In each panel the black open points are the observed data, while the gray filled points are our model predictions. The dashed line shows the \citetalias{blitz06b} empirical fit.
}
\end{figure*}

We can further check the level of agreement between the model and the data, and see how well our theory compares to the purely empirical fit, using the \citetalias{blitz06b} sample, where there is no (or less) systematic offset due to unresolved clumping. In Figure \ref{pres_single} we plot the same data as in Figure \ref{pres2}, but show the data and model predictions for the galaxies one by one. As the plots show, our model not only fits the general trend between $\rmol$ and $P$, for most galaxies we obtain a good match on a bin-by-bin basis. Even in those galaxies where the agreement is poor, our model still gives a correct lower limit for $\rmol$. The level of agreement between our predicted curves and the data is as good as the purely empirical fit between $\rmol$ and pressure in \citetalias{blitz06b}, without the need for any free parameters. It is not clear why one model or the other does better for particular galaxies. Our model appears to provide a substantially better fit for NGC3627, NGC5033, and NGC5055, while doing noticeably worse for IC 10, NGC0598, and NGC4736; for the rest of the sample the \citetalias{blitz06b} empirical fit and our model calculations are nearly equally good fits. It is not clear what galaxy properties favor one model or the other.

\subsubsection{The Small Magellanic Cloud}
\label{smcsec}

As noted above, the SMC's metallicity of $12+\log(\mbox{O}/\mbox{H})=8.0$ \citep{dufour84a} makes it the lowest metallicity galaxy in our sample for which we have detections rather than upper limits on the H$_2$ content. To analyze this data set, we take the maps of \hi and H$_2$ column determined by \citet{leroy07a} and break the data into 20 bins in total gas column (including He). In each bin we compute the mean \hi and H$_2$ column density. It is worth noting that the technique used to infer H$_2$ column densities in this data set have $\sim 50\%$ systematic uncertainties, so we must proceed with this caution in mind.

To test our theoretical model, we generate a prediction for the H$_2$ mass fraction as a function of total gas column density $\Sigmacomp$ as in equation (\ref{h2mfraceqn}), using a metallicity relative to Solar of $Z'\approx 0.2$. Before comparing the model curve to the data, we face the difficult problem of inclination correction. The SMC is a triaxial structure with an aspect ratio of roughly $1:2:4$, with $4$ representing the direction along the line of sight \citep{crowl01a}. The large extent along the line of sight means that the inclination correction is very significant, but the triaxiality of the galaxy means that no single number describes the inclination, as is the case for a disk. We therefore compare the model and the data using two different inclination corrections, one assuming an inclination $i=76^{\circ}$ (corresponding to a $1:4$ aspect ratio) and one assuming an inclination $i=63^{\circ}$ (corresponding to a $1:2$ aspect ratio). These should bracket the true inclination correction. Another caution we should make here is that, due to these large inclinations, there may be significant amounts of gas along a given line of sight that are not associated with whatever atomic-molecular complex it intersects, so the \hi column densities may be overestimates (see \S~\ref{sigmazdepend}).

\begin{figure}
\plotone{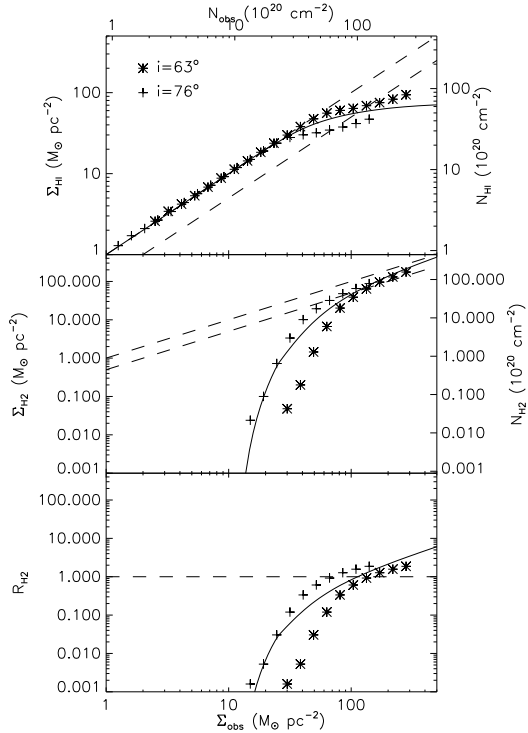}
\caption{
\label{smcplot}
Column densities of \hi and H$_2$ (\textit{upper two panels}) and H$_2$ to \hi ratio (\textit{bottom panel}) as a function of total column density $\Sigma_{\rm obs}$ in the SMC. The solid curve shows the model prediction, while the dashed lines show our confidence regions, as in Figures \ref{hicomp}--\ref{rmolcomp}. The asterisks and plus signs represent the observed column densities using inclination corrections of $63^{\circ}$ and $76^{\circ}$, respectively.
}
\end{figure}

Figure \ref{smcplot} shows a comparison of the data and the model curve. As expected, the data for the two different inclination corrections bracket the model curve quite well. We should be cautious about reading too much into the agreement at low $\Sigma_{\rm obs}$ ($\ltsim 30$ $\msun$ pc$^{-2}$), where both the data and our model are quite uncertain. Indeed, $\rmol$ is such a sharp function of $\Sigma_{\rm obs}$ at low column that the factor of $2$ variation in the total column introduced by the uncertainty in the inclination translates to a factor of $\sim 50$ difference in the value of $\rmol$ at a fixed $\Sigma_{\rm obs}$. Thus, our model need only be correct within a factor of 50 to lie in between the two curves for the different inclination corrections. At higher $\Sigma_{\rm obs}$ ($\gtsim 80$ $\msun$ pc$^{-2}$), on the other hand, the inclination correction only produces a factor of $\sim 2$ uncertainty in $\rmol$, and the ability of our model to match the data here is significant. It shows that we are capable of predicting the total column at which $\rmol\sim 1$ to better than a factor of 2 accuracy even in a galaxy with metallicity $\sim 1/10$ Solar.

\subsection{Galactic Observations}
\label{galobs}

\subsubsection{Data Sets}

We next compare our model to observations of the molecular content of clouds in the Milky Way, as measured by the \textit{Copernicus} \citep{spitzer75} and the \textit{Far Ultraviolet Spectroscopic Explorer (FUSE)} \citep{moos00, sahnow00} missions. These satellites measured absorption of ultraviolet light from background stars, and in some cases AGN, in the Lyman-Werner bands, enabling them to estimate the population of molecular hydrogen along a given line of sight.

\citet{savage77} and \citet{bohlin78} (for \textit{Copernicus}) and \citet{rachford02, rachford08a} (for \textit{FUSE}) report measurements of the molecular hydrogen columns in the Milky Way disk for lines of sight at low galactic latitude ($b\ltsim 10^{\circ}$), using stars as background sources. They combine these with \hi observations along the same lines of sight to determine molecular hydrogen fractions for clouds these sightlines. \citet{gillmon06a} report \textit{FUSE} observations of sightlines at high galactic latitude ($b\gtsim 20^{\circ}$) using AGN as background sources. These observations probe clouds lying above or below the galactic plane, which are presumably illuminated only on one side by stars in the disk. As with the low-latitude samples, \citeauthor{gillmon06a}\ combine these observations with measurements of the \hi column along the same sightlines to determine molecular fractions.

These observations are different than the extragalactic ones in that they are pencil-beam measurements rather than averages over atomic-molecular complexes. We possess little information about the geometry of the clouds these beams probe, and this ignorance complicates comparison of the data to our model. A given observed total (atomic plus molecular) column density $\Sigma_{\rm obs}$ might be the result of a line of sight passing through the center of a small cloud, or might be the result of a beam that passes only tangentially through a much larger cloud. Our models do not predict the same H$_2$ content in these two cases, so in the absence of additional information there is no way to map a given $\Sigma_{\rm obs}$ to a unique prediction for the molecular fraction. Similarly, there is good evidence in the data sets from the rotational excitation of the H$_2$ that at least some lines of sight probe multiple clouds that are separated in space \citep[e.g.][]{browning03}. Our model predicts a lower molecular fraction for this case than for the case of a single cloud with the same $\Sigma_{\rm obs}$. Attempts to remove these effects statistically are complicated by the fact that the \textit{FUSE} and \textit{Copernicus} lines of sight are not an unbiased sample. Each of the samples we use was chosen specifically to probe a certain range in column density or other properties, and all the selections are biased against high column densities and high molecular fractions, both of which produce high extinctions of LW photons that make determining a molecular column very costly or altogether impossible. Below we discuss how we deal with the problems of geometric uncertainty and selection bias for the galactic plane and high-latitude samples.

A second complication for this data set is that the assumption we make in our model that clouds are subjected to a relatively uniform background dissociating radiation field \citep[cf.][]{krumholz08c} may not be appropriate in this context. Treating the ISRF as uniform is reasonable for the giant cloud complexes with masses $\ga 10^4$ $\msun$ and sizes $\ga 100$ pc probed by extragalactic measurements; these sample the radiative output of many stars and star clusters. In contrast, the entire sample of high latitude clouds observed by \citet{gillmon06a} has a total mass $\sim 3000$ $\msun$ spread out over a $\sim 100$ pc$^2$ area \citep{gillmon06b}, making them tiny in comparison to giant complexes. The galactic clouds probed by the \textit{Copernicus} and \textit{FUSE} observations are of unknown size, but are likely to be similarly small. This is because any line of sight passing through a true giant cloud complex would be completely opaque in the Lyman-Werner bands, thus preventing \textit{Copernicus} or \textit{FUSE} from making any measurement of the H$_2$ column. The dissociating flux in the vicinity of such small clouds may be dominated by a single nearby star or star cluster.

Due to the effects of uncertain geometry and non-uniform radiation fields, we expect this data set to show significantly more scatter than the extragalactic one.

\subsubsection{The Galactic Plane Data}

Given the limitations of pencil-beam measurements, we must compare our model to the data in a way that accounts for our geometric uncertainty. The most straightforward comparison we can make is by considering a limiting case. Consider an observation that measures a total column density $\Sigmaobs$.  According to our model, the fraction of gas in molecular form will depend on the geometry of the complex; if the line of sight probes the outer parts of a very large cloud, always staying within the atomic region, then in principle there could be no molecules present no matter how large $\Sigmaobs$ might be. The reverse is not true, however. Even if the line of sight passes exactly through the center of an atomic-molecular complex, it must still pass through atomic shielding layer, and there will therefore be some \hi along that line of sight. In our model, therefore, there is a minimum amount of \hi that we predict must be present for a given column of H$_2$; this minimum corresponds to the case of a line of sight that passes directly through the center of a complex. This implies that we can make a prediction as follows: we consider a complex of mean column density $\Sigmacomp$, and use our model to predict the atomic, molecular, and total column densities, $\Sigma_{\rm HI,cen}$, $\Sigma_{\rm H_2,cen}$, and $\Sigma_{\rm cen}$, that one would see along a line of sight passing through the center of the complex. The curves $\Sigma_{\rm HI,cen}$ versus $\Sigma_{\rm cen}$ and $\Sigma_{\rm H_2,cen}$ versus $\Sigma_{\rm cen}$ that we generate through this procedure should then be lower and upper limits, respectively, on the observed distributions of $\Sigma_{\rm HI}$ versus $\Sigmaobs$ and $\Sigma_{\rm H_2}$ versus $\Sigmaobs$.

\begin{figure}
\plotone{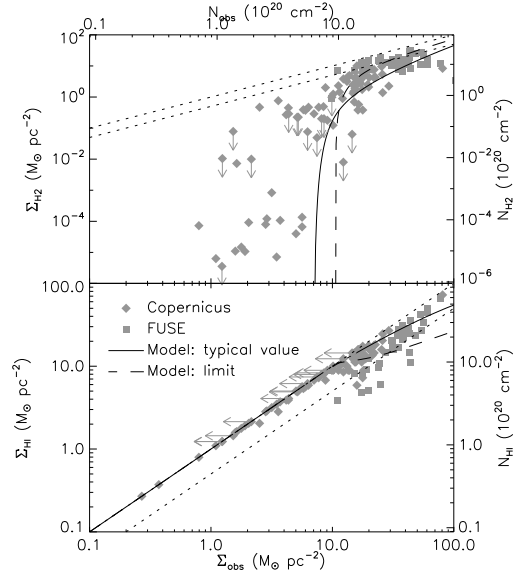}
\caption{\label{galcomp}
Column density of H$_2$ (\textit{upper panel}) and \hi (\textit{lower panel}) versus total column density for the \textit{Copernicus} (\textit{diamonds}) and \textit{FUSE} (\textit{squares}) data, and for the limits (\textit{dashed line}) and typical values (\textit{solid line}) we compute from our model. The dashed line is an upper limit in the upper panel, and a lower limit in the lower one. For the data, we only plot points with measured \hi columns. Arrows indicate lines of sight for which the \hi column is measured but only an upper limit is available for the H$_2$ column. In the upper panel the dotted lines show $\Sigma_{\rm H_2}=\Sigma$ and $\Sigma_{\rm H_2}=\Sigma/3$ (the edge of our region of confident prediction), while in the lower panel they show $\Sigma_{\rm HI}=\Sigma$ and $\Sigma_{\rm HI}=2 \Sigma/3$. Note that $N_{\rm H_2}$ represents the column of H nuclei in molecular form; the number of H$_2$ molecules is half this.
}
\end{figure}

Instead of computing a limit, we can also make an estimate for the ``typical" atomic and molecular fractions we should see at a given $\Sigmaobs$. If we knew the true distribution of column densities for atomic-molecular complexes in the Milky Way, we could do this by integrating all lines of sight through that distribution and computing the mean atomic and molecular columns at a given total column. However, we do not know the true distribution, and even if we did this procedure would not address effects of observational bias in selecting sightlines. Instead, we make a much rougher calculation. Using our model we can compute the radius $\xm(\Sigmacomp)$ at which the cloud whose true mean column density is $\Sigmacomp$ transitions from atomic to molecular for solar metallicity ($Z=1$). The atomic and molecular columns along a sightline that strikes the cloud of radius $R$ at a distance $\beta R$ from its center are
\begin{eqnarray}
\Sigma_{\rm HI} & = & 
\frac{2\tauR}{\sigmad}
\left\{
\begin{array}{ll}
\sqrt{1-\beta^2}-\sqrt{\xm^2-\beta^2}, \quad & \beta < \xm \\
\sqrt{1-\beta^2}, & \beta > \xm
\end{array}
\right. \\
\Sigma_{\rm H_2} & = & 
\frac{2\tauR\phimol}{\sigmad}
\left\{
\begin{array}{ll}
\sqrt{\xm^2-\beta^2}, \quad & \beta<\xm \\
0, & \beta > \xm
\end{array}
\right.,
\end{eqnarray}
and the H$_2$ fraction averaged over all pencil beams passing through the cloud is,
\begin{equation}
f_{\rm H_2,beams}(\Sigmacomp) =
2\int_0^{1} \beta \frac{\Sigma_{\rm H_2}}{\Sigma_{\rm H_2}+\Sigma_{\rm HI}}
d\beta.
\end{equation}
To estimate a ``typical" atomic or molecular content for a given observed total gas column $\Sigmaobs$, we can simply take $\Sigma_{\rm HI,obs} \approx [1-f_{\rm H_2,beams}(\Sigmaobs)]\Sigmaobs$ and $\Sigma_{\rm H2,obs} \approx f_{\rm H_2,beams}(\Sigmaobs)\Sigmaobs$. This is equivalent to saying that our sightlines do indeed probe random impact parameters on cloud complexes, and that when we observe a sightline of column density $\Sigmaobs$, most of the time we are observing a complex whose mean column density is also about $\Sigmaobs$. This assumption could fail if low column density clouds were rare compared to high column density ones, so that most of the sightlines that produce a given $\Sigmaobs$ were really tangential paths through high column clouds rather than beams passing close to the the centers of low column clouds. However, there is no evidence that high-column density atomic-molecular complexes outnumber low-column ones, and in general in the ISM more mass tends to be in diffuse than dense structures, so we proceed with our assumption that $\Sigmacomp\approx \Sigmaobs$. 

We plot both our limits on the \hi and H$_2$ columns and our estimates for their typical values against the \textit{Copernicus} and \textit{FUSE} data sets in Figure \ref{galcomp}. As the plot shows, both our limits and our typical columns match the observed data reasonably well. As with the extragalactic case, we recover the overall trend that the gas is mostly \hi until a total column of a few $\msun$ pc$^{-2}$, and then mostly molecular thereafter.

\subsubsection{The High-Latitude Data}

\begin{figure}
\plotone{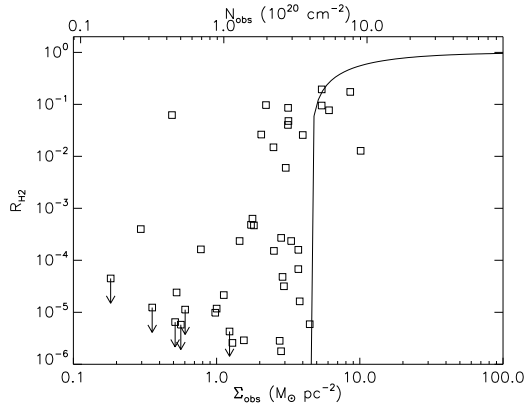}
\caption{\label{hlcomp}
$\rmol$ versus column density for the \textit{FUSE} high-latitude sample (\textit{squares}; \citealt{gillmon06a}) and for our model (\textit{line}). Arrows indicate lines of sight for which only an upper limit is available for the H$_2$ column. Note that $N_{\rm obs}$ represents the total column of H nuclei in either atomic or molecular form.
}
\end{figure}

The comparison to the high-latitude data set is somewhat less straightforward, because the clouds at high latitude are only illuminated from one side. Indeed, \citet{gillmon06a} find that these clouds become molecular at column densities a factor of $\sim 2$ lower than do clouds in the galactic plane, probably as a result of this one-sided illumination. We have not considered the case of clouds of finite size illuminated over only half their surfaces. However, we can obtain a reasonable approximation to this configuration using the case of \textit{semi-infinite} clouds subjected to an isotropic, uniform dissociation radiation field, which we considered in \S~\ref{giantclouds}. A semi-infinite cloud blocks dissociating radiation over $2\pi$ sr, and thus the depth of the atomic shielding layer at its surface is determined only by photons that arrive from the ``front side". This is therefore a close analogy to the case of a high-latitude cloud, although it differs in the cloud geometry.

In Paper I we show that using a semi-infinite slab in place of a finite cloud tends to produce errors in estimating the exact location of a transition from atomic to molecular in a finite cloud, but that the semi-infinite calculation does give a good estimate of the column at which clouds transition from mostly atomic to mostly molecular. We therefore expect to obtain roughly the right transition column and thus the right column of \hi for clouds that are mostly molecular, but not particularly accurate predictions for the exact molecular column. For a semi-infinite cloud of solar metallicity, as we show in \S~\ref{giantclouds} we predict a molecular to atomic ratio of
\begin{equation}
\rmol =
\left\{
\begin{array}{ll}
0, & \Sigma < 4.5\,\msun\mbox{ pc}^{-2} \\
(\Sigma-4.5\,\msun\mbox{ pc}^{-2})/\Sigma, & \Sigma > 4.5\,\msun\mbox{ pc}^{-2}
\end{array}
\right.,
\end{equation}
where $\Sigma$ is the total column density. In other words, for a semi-infinite cloud of solar metallicity, the first $4.5$ $\msun$ pc$^{-2}$ are atomic, and the rest are molecular. We show this prediction overplotted with the \textit{FUSE} high-latititude data in Figure \ref{hlcomp}. As the plot shows, we indeed do not get very accurate predictions for the exact ratio of \hi to H$_2$, but our calculation agrees quite well with the general value of $\Sigma$ for which the transition from molecular to atomic occurs. In particular, our one-sided model recovers the observational result that a cloud illuminated from one side shows a lower atomic-to-molecular transition column than an isotropically-illuminated cloud.

\section{Summary and Conclusions}
\label{conclusion}

We present a first-principles calculation of the molecular gas content of galactic disks in terms of the observable properties of those galaxies. Our calculations build on the simple model for photodissociation fronts in finite clouds presented in \citetalias{krumholz08c}, in which we show that the amount of atomic material required to shield a molecular cloud against dissociation by the interstellar radiation field (ISRF) can be characterized by two parameters: $\chi$, a radiation field strength normalized by the gas density and the properties of dust grains, and $\tauR$, a measure of the dust optical depth of a cloud. We show that, due to the way the density in the cold phase of the atomic ISM varies with ISRF, the normalized radiation field strength takes a characteristic value $\chi\approx 1$ in all galaxies where a two-phase atomic ISM is present, with only a weak dependence on metallicity. 

The existence of a characteristic normalized radiation field strength, and its weak dependence on metallicity, has a number of important consequences. First, it enables us to give a simple analytic approximation (equations \ref{hifracanalyt} or \ref{hifracanalyt1}) for the fraction of mass in an atomic-molecular complex that will be in the atomic or molecular phases solely in terms of the column density of the complex and the metallicity of the gas. This makes it easy to test our calculations against observations. Second, we show that as a consequence of $\chi$ assuming a nearly fixed value, the atomic envelopes of molecular clouds have a characteristic visual extinction $A_V\approx 0.2$ at solar metallicity. Similarly, the transition from atomic to molecular gas occurs at a characteristic shielding column of $\Sigma_{\rm HI} \approx 10$ $\msun$ pc$^{-2}$. These quantities both vary sub-linearly metallicity, with $A_V$ declining and $\Sigma_{\rm HI}$ increasing as metallicity does. We calculate these values and their metallicity-dependence solely in terms of the microphysical constants that describe the properties of H$_2$ formation and dissociation, the cooling curves of C\textsc{ii} and O\textsc{i}, and the properties of interstellar dust grains. Our model does not depend on unobservable parameters such as the gas volume density or the ISRF strength in a galaxy.

Our model compares favorably with observations of the atomic and molecular content of clouds both in the Milky Way and in nearby galaxies. We are able to reproduce both the characteristic column density at which clouds transition from being primarily atomic to primarily molecular, and the way that this characteristic column depends on metallicity and on whether clouds are illuminated on one side or on both sides. We are also able to reproduce the observed correlation between molecular content and interstellar pressure.

The development of a predictive model for the molecular content of galaxies that does not rely on unknown and generally unobservable quantities such as the radiation fields or gas volume densities in those galaxies opens up new possibilities to advance our understanding of galactic evolution. In low-density dwarfs or high-redshift galaxies containing few metals, the formation of molecular clouds may be the rate limiting step in star formation. On the other hand, previous work has shown that, once molecular gas forms, it converts itself into stars at a rate of a few percent of the mass per free-fall time independent of its density or environment \citep{krumholz05c, krumholz07e, mckee07b}. Thus a theory of molecule formation creates the possibility of developing a theory of the star formation rate capable of making predictions that can be applied not only in the relatively molecule-rich nearby galaxies for which empirical star formation laws have been determined \citep[e.g.][]{kennicutt98a}, but also in the more distant and lower metallicity universe where these laws are known to break down \citep[e.g.][]{wolfe06a}. We plan to explore such a theory in future work.

\acknowledgements 
We thank B.~Rachford and A.~Leroy for providing copies of their data, and F.~Bigiel, S.~Faber, N.~Gnedin, A.~Leroy, A.~Sternberg, and F.~Walter for helpful discussions and comments on the manuscript. We thank the referee, E.~Rosolowsky, for providing comments that improved the quality of the paper. Support
for this work was provided by: NASA through Hubble Fellowship grant
\#HSF-HF-01186 awarded by the Space Telescope Science Institute, which
is operated by the Association of Universities for Research in
Astronomy, Inc., for NASA, under contract NAS 5-26555 (MRK); NASA, as part of the Spitzer Space Telescope Theoretical Research Program, through a contract issued by the Jet Propulsion Laboratory, California Institute of Technology (MRK); and by the National Science Foundation through grants AST-0807739 (to MRK), AST-0606831 (to CFM), and PHY05-51164 (to the Kavli Institute for Theoretical Physics, where MRK, CFM, and JT collaborated on this work).


\end{document}